\documentclass[a4paper]{article}
\usepackage[margin=1in]{geometry} % full-width
\usepackage[fleqn]{amsmath}
\usepackage{amsthm}
\usepackage{amssymb}

\usepackage{mathtools}
\mathchardef\mhyphen="2D

\usepackage[nolist,nohyperlinks]{acronym}
\acrodef{2DMatPedia}{2D Materials Encyclopedia}
\acrodef{AFLOW}{Automatic Flow}
\acrodef{AiiDA}{Automated Interactive Infrastructure and Database for Computational Science}
\acrodef{ALIGNN}{Atomistic LIne Graph Neural Network}
\acrodef{ASE}{Atomic Simulation Environment}
\acrodef{AUC}{Area Under the Curve}
\acrodef{AutoML}{Automated Machine Learning}
\acrodef{C2DB}{Computational 2D Materials Database}
\acrodef{CGCNN}{Crystal Graph Convolutional Neural Network}
\acrodef{CLI}{Command-Line Interface}
\acrodef{CMR}{Computational Materials Repository}
\acrodef{CN}{Coordination Number}
\acrodef{DFT}{Density-Functional Theory}
\acrodef{DoE}{Department of Energy}
\acrodef{FAIR}{Findable, Accessible, Interoperable, and Reusable}
\acrodef{FPR}{False Positive Rate}
\acrodef{FWE}{Family-wise Error}
\acrodef{GA}{Genetic Algorithm}
\acrodef{GCN}{Generalized Coordination Number}
\acrodef{GII}{Global Instability Index}
\acrodef{GNN}{Graph Neural Network}
\acrodef{HT}{High-Throughput}
\acrodef{HHI}{Herfindahl-Hirschman Index}
\acrodef{JARVIS}{Joint Automated Repository for Various Integrated Simulations}
\acrodef{LASSO}{Least Absolute Shrinkage and Selection Operator}
\acrodef{MAE}{Mean Absolute Error}
\acrodef{MEGNet}{Materials Graph Network}
\acrodef{MGI}{Materials Genome Initiative}
\acrodef{MOF}{Metal-Organic Framework}
\acrodef{ML}{Machine Learning}
\acrodef{MSE}{Mean Squared Error}
\acrodef{NAS}{Neural Architecture Search}
\acrodef{NN}{Neural Network}
\acrodef{NOMAD}{Novel Materials Discovery}
\acrodef{OQMD}{Open Quantum Materials Database}
\acrodef{ROC}{Receiver Operating Characteristic}
\acrodef{Roost}{Representation Learning from Stoichiometry}
\acrodef{RMSE}{Root Mean Squared Error}
\acrodef{SBIR}{Small Business Innovation Research}
\acrodef{SMOTE}{Synthetic Minority Oversampling Technique}
\acrodef{SISSO}{Sure Independence Screening and Sparsifying Operator}
\acrodef{TPE}{Tree-structured Parzen Estimator}
\acrodef{TPOT}{Tree-based Pipeline Optimization Tool}
\acrodef{TPR}{True Positive Rate}
\acrodef{UFF}{Universal Force Field}
\acrodef{XGBoost}{eXtreme Gradient Boosting}

\usepackage{booktabs}
\usepackage[table]{xcolor}
\usepackage{multirow}

\usepackage[utf8]{inputenc}
\usepackage{hyperref}
\hypersetup{
	unicode,
	pdfauthor={James Dean, Matthias Scheffler, Tom Purcell, Sergey V Barabash, Rahul Bhowmik, Timur Bazhirov},
	pdftitle={Interpretable Machine Learning for Materials Design},
	pdfsubject={Comparison of various machine learning models with respect to accuracy and interpretability in materials design.},
	pdfkeywords={machine learning, materials science, chemistry, interpretability, inverse design},
	pdfproducer={LaTeX},
	pdfcreator={pdflatex}
}

\usepackage[sort&compress,numbers,square,super]{natbib}
\usepackage{graphicx}
\usepackage[labelfont=bf, font=footnotesize]{caption}
\usepackage{float}
\usepackage{indentfirst}

\usepackage{listings}
\usepackage{xcolor}
\definecolor{codegreen}{rgb}{0,0.6,0}
\definecolor{codedarkblue}{rgb}{0.039,0.188,0.411}
\definecolor{codegray}{rgb}{0.5,0.5,0.5}
\definecolor{codepurple}{rgb}{0.58,0,0.82}
\definecolor{backcolour}{rgb}{0.95,0.95,0.92}
\definecolor{backcolourGray}{rgb}{0.97,0.97,0.97}

\lstdefinestyle{mystyle}{
    backgroundcolor=\color{backcolourGray},   
    commentstyle=\color{codegreen},
    keywordstyle=\color{magenta},
    numberstyle=\tiny\color{codegray},
    stringstyle=\color{codedarkblue},
    basicstyle=\footnotesize\ttfamily,
    breakatwhitespace=false,         
    breaklines=true,                 
    captionpos=b,                    
    keepspaces=true,
    numbers=left,                    
    numbersep=2pt,                  
    showspaces=false,                
    showstringspaces=false,
    showtabs=false,                  
    tabsize=2
}

\usepackage{comment}
\usepackage{todonotes}
\usepackage{authblk}

\usepackage[table]{xcolor}

\title{\textbf{Interpretable Machine Learning for \\ Materials Design}}
\author[1]{James Dean}
\author[2,3]{Matthias Scheffler}
\author[3]{Thomas A. R. Purcell}
\author[4]{Sergey V. Barabash}
\author[5]{Rahul Bhowmik}
\author[1,*]{Timur Bazhirov}

\affil[1]{Exabyte Inc, San Francisco, CA, United States}
\affil[2]{University of California Santa Barbara, Isla Vista, CA, United States}
\affil[3]{The NOMAD Laboratory at the Fritz Haber Institute, Berlin, Germany}
\affil[4]{Intermolecular Inc, San Jose, CA, United States}
\affil[5]{Polaron Analytics, Beavercreek, OH, United States}
\affil[*]{Corresponding Author Email: \texttt{timur@exabyte.io}}

\date{\today}

\begin{document}
	\maketitle

% \listoftodos
% \TP{Test TP}

% \JD{Test JD}

% \MS{Test MS}

% \SB{Test SB}

% \RB{Test RB}

% \TB{Test TB}

\begin{abstract}

    Fueled by the widespread adoption of \acl{ML} and the high-throughput screening of materials, the data-centric approach to materials design has asserted itself as a robust and powerful tool for the \textit{in-silico} prediction of materials properties.
    %\todo{science was always "data driven" therefore I prefer "data-centric": Science is and always has been based on data, but the term data-centric indicates a radical shift in the way information is handled and research is performed. It refers to extensive data collections, digital repositories, and new concepts and methods of data analytics. It also implies that we complement the traditional purpose-oriented research by using data from other studies.}
    % Makes sense - I agree, "data-centric" does convey the current shift in how research and data interact a lot better than "data-driven" does. Changed the wording as a result. - JD
    When training models to predict material properties, researchers often face a difficult choice between a model's interpretability or its performance. We study this trade-off by leveraging four different state-of-the-art \acl{ML} techniques: \acs{XGBoost}, \acs{SISSO}, \acs{Roost}, and \acs{TPOT} for the prediction of structural and electronic properties of perovskites and 2D materials. We then assess the future outlook of the continued integration of \acl{ML} into materials discovery, and identify key problems that will continue to challenge researchers as the size of the literature's datasets and complexity of models increases. Finally, we offer several possible solutions to these challenges with a focus on retaining interpretability, and share our thoughts on magnifying the impact of \acl{ML} on materials design.
    		
\end{abstract}

\section{Introduction}
\label{sec:introduction}
    
    Today, big data and artificial intelligence revolutionize many areas of our daily life, and materials science is no exception\cite{draxlBigDataDrivenMaterials2020, materDeepLearningChemistry2019, butlerMachineLearningMolecular2018}.
    % \todo{a significant part of the Schleder et al paper was stolen from a preprint (2018) when one of the coauthors was a postdic. I prefer to cite our paper: Draxl\&Scheffler in Handbook ... 2020.}.
    % Didn't realize part of that work was taken from another work. I've replaced it with the Draxl/Scheffler work that appears in the following Handbook of Materials Modeling: https://link.springer.com/referencework/10.1007%2F978-3-319-44677-6. Let me know if it's not the right one. -JD
    % \todo{I am not so happy with "advent" and with "has accelerated" because we have not seen (much) bigger success. I would probably start with: Today, big data and artificial intelligence revolutionize many areas of our daily life, and materials science is no exception.}
    % Makes sense; I've updated the paragraph to begin with ths. - JD
    More scientific data is available now than ever before and the size of the literature is growing at an exponential rate \cite{bornmannGrowthRatesModern2015, priceLittleScienceBig1963, priceScienceBabylon1975, priceNetworksScientificPapers1965}. This has led to multiple efforts in building the digital ecosystem for material discovery, most notably the \ac{MGI}
    \cite{nationalscienceandtechnologycouncilMaterialsGenomeInitiative2011, subcommitteeonthematerialsgenomeinitiativecommitteeontechnologyMaterialsGenomeInitiative2021}.
    % \todo{why don't we cite the original Whitehouse papers from 2011 and the one from just now, 2021. They are quite good.}.
    % Sure - swapped these out with the White House OSTP report from 2011, and the MGI's strategic plan from 2021. Couldn't find a press release direct from the White House this year; let me know if there's one in particular that would be better. -JD
    The \ac{MGI} is a multinational effort focused on improving the tools and techniques surrounding materials research, which recently has included suggestions to adopt the set of \ac{FAIR} principles when reporting data \cite{depabloNewFrontiersMaterials2019}. In the years since the creation of the \ac{MGI}, a number of large materials and chemical datasets have emerged, including
    the \ac{2DMatPedia} \cite{zhou2DMatPediaOpenComputational2019},
    \ac{AFLOW} database  \cite{curtaroloAFLOWAutomaticFramework2012, curtaroloAFLOWLIBORGDistributed2012},
    \ac{C2DB}\cite{gjerdingRecentProgressComputational2021, haastrupComputational2DMaterials2018},
    \ac{CMR}\cite{landisComputationalMaterialsRepository2012},
    \ac{JARVIS}\cite{choudharyJointAutomatedRepository2020},
    Materials Project \cite{jainCommentaryMaterialsProject2013},
    \ac{NOMAD} repository\cite{draxlNOMADLaboratoryData2019},
    and the \ac{OQMD}\cite{kirklinOpenQuantumMaterials2015}. We note that all of these are primarily computational in nature, and that there is still a scarcity of large databases containing comprehensively-characterized experimental data. Despite this, at least in computational materials discovery, the current availability of data has been a boon for exploration of the materials space, as it allows for highly flexible, data-hungry\cite{vanderploegModernModellingTechniques2014} models to be trained.
    % \todo{Shouldn't we tell that this is all computational, so far. We still have a problem that "comprehensively characterized experimental data" are very few.}
    % Changed the wording to make it clearer that these datasets are computational, and that we're still lacking in large experimental datasets. -JD
    
    One such approach that has seen widespread popularity in recent years is gradient boosting. Gradient boosting \cite{masonBoostingAlgorithmsGradient2000} is an ensemble technique in which a collection of weak learners (typically decision trees) are incrementally trained with respect to the gradient of the loss function \cite{hastieElementsStatisticalLearning2009}. A well-known implementation (with over 5,500 citations as of November 2021) is \ac{XGBoost}\cite{chenXGBoostScalableTree2016}, which reformulates the algorithm to provide stronger regularization and improved protection against over-fitting. In chemistry, its applications have been diverse: \ac{XGBoost} has been used to predict the adsorption energy of noble gases to \acp{MOF}\cite{liangXGBoostOptimalMachine2021}, biological activity of pharmaceuticals\cite{husnaComparisonMachineLearning2020}, atmospheric transport\cite{ivattImprovingPredictionAtmospheric2020}, and has even been combined with the representations found in \acp{GNN} to generate accurate models of various molecular properties \cite{dengXGraphBoostExtractingGraph2021}.
    
    Neural networks have also seen a lot of interest, owing to their greater flexibility relative to other model types. This has included
    % \todo{is this true? I would start with the second part (decission trees)}
    % Moved the section on decision trees / xgboost to appear directly after the database segment. -JD
    the influential Behler-Parinello \cite{behlerGeneralizedNeuralNetworkRepresentation2007} and \ac{CGCNN}\cite{xieCrystalGraphConvolutional2018} architectures based on chemical structure, the \ac{Roost}\cite{goodallPredictingMaterialsProperties2020} architecture based on chemical formula, and many other approaches\cite{behlerFourGenerationsHighDimensional2021, yaoTensorMol0ModelChemistry2018, westermayrCombiningSchNetSHARC2020, schuttSchNetContinuousfilterConvolutional2017, toniatoUnassistedNoiseReduction2021, schwallerPredictingRetrosyntheticPathways2020, schwallerMolecularTransformerModel2019, schwallerFoundTranslationPredicting2018, vaucherInferringExperimentalProcedures2021, materDeepLearningChemistry2019, panteleevRecentApplicationsMachine2018, draxlBigDataDrivenMaterials2020}. 
    
    The modern \ac{ML} toolbox is large, although it is still far from complete.
    % \todo{it is still far from complete}
    % Added this statement - JD
    As a result model selection techniques are becoming increasingly necessary: this has led to the field of \ac{AutoML}. This area of work has seen much progress in recent years \cite{gijsbersOpenSourceAutoML2019, yaoTakingHumanOut2019}, and has even been extended to \ac{NAS}\cite{heAutoMLSurveyStateoftheart2021}, or the automated optimization of neural network architectures. In this work, we leverage the \ac{TPOT} approach to \ac{AutoML}\cite{leScalingTreebasedAutomated2020, olsonAutomatingBiomedicalData2016, olsonEvaluationTreebasedPipeline2016}, which uses a \ac{GA} to create effective \ac{ML} pipelines. Although it generally draws from the models of SciKit-Learn \cite{scikit-learn}, it can also be configured to explore gradient boosting models via \ac{XGBoost} \cite{masonBoostingAlgorithmsGradient2000}, and neural network models via PyTorch \cite{NEURIPS2019_9015}. Moreover, \ac{TPOT} also performs its own hyperparameter optimization, thus providing a more hands-off solution to identifying \ac{ML} pipelines. The success of \ac{GA}-based approaches in \ac{ML} is not isolated to \ac{AutoML}. Indeed, they are a fundamental part of genetic programming, where they are used to optimize functions for a particular task\cite{amirhaeriStatisticalGeneticProgramming2017, kinnearAdvancesGeneticProgramming1994}. Eureqa\cite{schmidtDistillingFreeFormNatural2009} is a particularly successful example of this\cite{stoutemyerCanEureqaSymbolic2012}, leveraging a \ac{GA} to generate equations fitting arbitrary functions, and has been used in several areas of chemistry, including the generation of adsorption models to nanoparticles\cite{deanUnfoldingAdsorptionMetal2019} and metal atoms to oxide surfaces\cite{tanPredictingMetalSupport2019}. This approach of fitting arbitrary functions to a task is also known as ``symbolic regression.'' Recent work surrounding compressed sensing has yielded the \ac{SISSO} approach\cite{ouyangSISSOCompressedsensingMethod2018}. \ac{SISSO} also generates equations mapping descriptors to a target property, proceeding by combining descriptors using various building blocks, including trigonometric functions, logarithms, addition, multiplication, exponentiation, and many others. This methodology has been highly successful in a variety of areas including crystal structure classification \cite{ouyangSimultaneousLearningSeveral2019}, as well as the prediction of perovskite properties \cite{ihalageAnalogicalDiscoveryDisordered2021, bartelNewToleranceFactor2019, xieMachineLearningOctahedral2020} and 2D topological insulators\cite{acostaAnalysisTopologicalTransitions2018}.
    
    The abundance of scientific data in the literature increasingly makes the use of highly flexible (yet data-hungry) techniques tractable. Although such techniques may deliver models which are highly accurate and generalize well to new data, what is oftentimes lost is the physical interpretation of these models. By physical interpretation, we mean an understanding of the relationship between the chosen descriptors and the target property. Although a black-box model which has a high level of accuracy but little physical interpretation may lend itself well to the Edisonian screening of a wide range of materials, it may be difficult to understand exactly what feature (or combination of features) actually matters to the design of the material. Once the screening is done and the target values are calculated, little may be done to improve performance aside from including new features, adjusting the model's hyperparameters, or increasing the size of the training set. Alternatively, consider a model which has less accuracy, but which has an intuitive explanation, such as an equation describing an approximate relationship between features and target. Although such a model may at first glance seem less useful than a highly-accurate black-box, such a model can help deliver insight into the underlying process that results in the target property. Moreover, by understanding which features are important, the model can give clues into what may be done to further improve it --- driving the rational discovery of materials. In addition, interpretability versus accuracy is not a strict trade-off,
    % \todo{I would no (not yet) say often. This would also contradict waht we say in the abstract.} 
    % Rephrased to remove that language - JD
    and it is possible for interpretable and black-box models to deliver similar accuracy\cite{rudinStopExplainingBlack2019}. Therefore, in this work we take steps to compare the performance of all four models for each problem with respect to i) performance and ii) interpretability.
    
    We leverage a diverse selection of techniques in order to draw comparisons of model accuracy and interpretability. Taking advantage of the current abundance of chemical data, we can re-use the \ac{DFT} calculations of others stored on several \ac{FAIR} chemical datasets. A series of three problems are investigated: 1) the prediction of perovskite volumes, 2) the prediction of 2D material bandgaps, and 3) the prediction of 2D material exfoliation energies. For the perovskite volume problem, we leverage the ABX\textsubscript{3} perovskite dataset (containing 144 examples) published by Körbel, Marques, and Botti\cite{korbelStabilityElectronicProperties2016}; this dataset is hosted by \ac{NOMAD}\cite{draxlNOMADLaboratoryData2019}, whose repository has strong focus on enabling researchers to report their data such that it satisfies the \ac{FAIR} data principles\cite{draxlNOMADFAIRConcept2018}. For the 2D material problems, we apply the \ac{2DMatPedia} published by Zhou et al\cite{zhou2DMatPediaOpenComputational2019}; this is a large dataset of 6,351 hypothetical 2D materials identified via a high-throughput screening of systems on the Materials Project\cite{jainCommentaryMaterialsProject2013}. We are aware of other 2D material datasets, such as the \ac{C2DB}\cite{gjerdingRecentProgressComputational2021} and \ac{JARVIS}\cite{choudharyJointAutomatedRepository2020}, but for the purposes of simplifying our study we only choose one. 
    
    We find that \ac{TPOT} delivers high-quality models, generally outperforming the other methods in terms of fitness metrics. Despite this, interpretability is not guaranteed, as it can create highly complex pipelines. \ac{XGBoost} lends itself to interpretation more consistently, as it allows for an importance metric, although it may be harder to understand exactly what the relationship is between the different features (or combinations of different features) and the target variable. We found that \ac{Roost} performed well on problems that could be approached via compositional descriptors (i.e. without structural descriptors); as a result, it can help us understand when a target property requires more than just the composition. Finally, we achieve the easiest interpretability from \ac{SISSO}, as it provides access to descriptors which directly capture the relationship between the features and target variable. Using these results, we discuss the advantages and disadvantages of each method, and discuss areas where the digital ecosystem surrounding materials discovery could be improved to improve adherence to \ac{FAIR} principles. Our work provides a comparison of several common \ac{ML} techniques on challenging (but relevant) materials property prediction problems.
    
  The manuscript is organized as follows: we begin by training a diverse set of four models, which are \ac{XGBoost}, \ac{TPOT}, \ac{Roost}, and \ac{SISSO} to investigate each of the three problems, resulting in a total of 24 trained models. Performance metrics (and comparative plots) are presented for each trained model to facilitate comparison, and we discuss the interpretation we can achieve from each of these models. Finally, we provide a discussion of the future outlook of \ac{ML} in the digital materials science ecosystem and what can be done to further accelerate materials discovery.

\section{Methodology}
\label{sec:methodology}

    \subsection{Data Sources}
    \label{methodology:data-sources}
        Crystal structures for the perovskite systems were obtained from the ``Stable Inorganic Perovskites'' dataset published by Körbel, Marques, and Botti
        % \todo{check: often one lists all authors when there are less than or equal to 3.}
        % Thanks for catching this - updated accordingly. -JD
        \cite{korbelStabilityElectronicProperties2016}, as hosted by \ac{NOMAD} \cite{draxlNOMADLaboratoryData2019}. This dataset contains a total of 144 \ac{DFT}-relaxed inorganic perovskites identified via a high-throughput screening strategy. Using this dataset, we develop a model of perovskite volume. As we rely on the use of compositional descriptors for these systems, we have scaled the volume of the perovskite unit cell by the number of formula units, such that the volume has units of Å\textsuperscript{3} / formula unit.
        
        Structures for 2D materials were obtained from the \ac{2DMatPedia} \cite{zhou2DMatPediaOpenComputational2019}, a large database containing a mixture of 6,351 real and hypothetical 2D systems. This database was generated via a \ac{DFT}-based high-throughput screening approach, which investigated bulk structures hosted by the Materials Project database \cite{jainCommentaryMaterialsProject2013} to find systems which may plausibly form 2D structures. Among other things, the \ac{2DMatPedia} provides \ac{DFT}-calculated exfoliation energies and bandgaps, along with a \ac{DFT}-optimized structure for each material. We use this dataset to develop models for the bandgap and exfoliation energy of 2D materials. Although the dataset reports exfoliation energies in units of eV, we have converted these into units of J / m\textsuperscript{2}. Bandgaps are reported in units of eV.
    
    \subsection{Feature Engineering}
    \label{methodology:feature-engineering}
        To facilitate the development of \ac{ML} algorithms capable of rapidly predicting material properties, we focus primarily on features that do not require further (computationally-intensive) \ac{DFT} calculations. In the case of the 2D material bandgap, we include the \ac{DFT}-calculated bandgap of the respective bulk material; we note that these values are tabulated on the Materials Project and can be looked up, thus circumventing the need for further \ac{DFT} work. A variety of chemical featurization libraries were used to generate compositional and structural descriptors for the systems we investigated, and are listed in sections \ref{methodology:feature-engineering:compositional-descriptors} and \ref{methodology:feature-engineering:structural-descriptors:matminer-descriptors} respectively. Features with values of NaN (which occurred when a feature could not be calculated) were assigned a value of 0. 
        % \todo{consider moving this line to the SI}
        % There wasn't a good place to put this in the SI; wasn't sure if it would be worth opening a new section just for this line. That said, I can move it if it strongly undercuts our paper. -JD
        
        \subsubsection{Compositional Descriptors}
        \label{methodology:feature-engineering:compositional-descriptors}
            Compositional (i.e. chemical formula-based) descriptors were calculated via the open-source XenonPy packaged developed by Yamada et al \cite{yamadaPredictingMaterialsProperties2019}. XenonPy uses tabulated elemental data from Mendeleev \cite{mentelMendeleevPythonResource2014}, Pymatgen \cite{ongPythonMaterialsGenomics2013}, the CRC Handbook of Chemistry and Physics \cite{rumbleCRCHandbookChemistry2021}, and Magpie \cite{wardGeneralpurposeMachineLearning2016} in order to calculate compositional features. XenonPy does this by combining the elemental descriptors (e.g. atomic weight, ionization potential, etc.) in various ways to form a single composition-weighted value. For example, three compositional descriptors may be obtained with XenonPy by taking the composition-weighted average, sum, or maximum elemental value of the atomic weight. Leveraging the full list of compositional features implemented in XenonPy results to 290 compositional descriptors, which are explained in greater detail within their publication \cite{yamadaPredictingMaterialsProperties2019}. 
            
            The compositional descriptors were used for the perovskite volume prediction, 2D material bandgap, and 2D material exfoliation energy prediction problems.
        
        \subsubsection{Structural Descriptors}
        \label{methodology:feature-engineering:structural-descriptors:matminer-descriptors}
            Some structural descriptors were calculated using MatMiner \cite{wardMatminerOpenSource2018}, an open-source Python package geared towards data-mining material properties. Leveraging MatMiner, the following descriptors were calculated: Average bond length, average bond angle, \ac{GII} \cite{SALINASSANCHEZ1992201}, Ewald Summation Energy \cite{ewaldBerechnungOptischerUnd1921}, a Shannon Information Entropy-based Structural Complexity (both per atom and per cell), and the number of symmetry operations available to the system. In the case of the average bond length and average bond angle, bonds were determined using Pymatgen's implementation of the JMol \cite{jmol} AutoBond algorithm. This list of bonds was also used to calculate an average \ac{CN} over all atoms in the unit cell. Finally, we also took the perimeter:area ratio of the 2D material's repeating unit.
            
            The structural descriptors were used for the 2D material bandgap and 2D material exfoliation energy problems.
        
    \subsection{Data Filtering}
    \label{methodology:data-filtering}
        The choice of data filtering methodology was chosen based on the problem at-hand. The perovskite volume prediction problem did not utilize any data filtering. In the case of the 2D material bandgap and exfoliation energy prediction problems, the data obtained from the \ac{2DMatPedia} were required to satisfy all of the following criteria:
        \begin{enumerate}
            \item No elements from the f-block, larger than U, or noble gases were allowed.
            \item Decomposition energy must be below 0.5 eV/atom.
            \item Exfoliation energy must be strictly positive.
        \end{enumerate}
        Additionally, in the case of the 2D material bandgap, data were required to have a parent material defined on the Materials Project. This was done because we use the Materials Project's tabulated \ac{DFT} bandgap of the bulk system as a descriptor for the bandgap of the corresponding 2D system.

    \subsection{ML Models}
    When training models, 10\% of randomly selected data was held out as a testing set. The same train/test split was used for all 4 models considered (\ac{XGBoost}, \ac{TPOT}, \ac{Roost}, and \ac{SISSO}). To facilitate a transparent comparison between models, in all cases we report the \ac{MAE}, \ac{RMSE}, Maximum Error, and R\textsuperscript{2} score of the test set.

    \label{methodology:ml-models}
        \subsubsection{Gradient-Boosting with XGBoost}
        \label{methodology:ml-models:gradient-boosting-with-xgboost}
            When training \ac{XGBoost} models, 20\% of randomly selected data was held out as an internal validation set. This was used to adopt an early-stopping strategy, where if the model \ac{RMSE} did not improve after 50 consecutive rounds, training was halted early. When training, \ac{XGBoost} was configured to optimize its \ac{RMSE}.
        
            Hyperparameters were optimized via the open-source Optuna \cite{akibaOptunaNextgenerationHyperparameter2019} framework. The hyperparameter space was sampled using the \ac{TPE} approach \cite{bergstraAlgorithmsHyperParameterOptimization2011, bergstraMakingScienceModel2013}. To accelerate the hyperparameter search, we leveraged the Hyperband \cite{liHyperbandNovelBanditbased2017} approach for model pruning, using the validation set \ac{RMSE} to determine whether to prune a model. Hyperband's budget for the number of trees in the ensemble was set to range between 1 and 256 (corresponding with the maximum number of estimators we allowed an \ac{XGBoost} model to have). The search space for hyperparameters is found in Table \ref{tab:methodology:ml-models:gradient-boosting-with-xgboost:hyperparameters}. 
            
             \begin{table}[H]
                    \centering
                    \rowcolors{2}{gray!25}{white}
                    \caption{Ranges of hyperparameters screened with Optuna for all \ac{XGBoost} runs. The search was inclusive of the listed minima and maxima. Hyperparameters use the same variable naming convention as in the \ac{XGBoost} documentation.}
                    \begin{tabular}{rcc}
                        \rowcolor{gray!50}
                        Hyperparameter & Minimum & Maximum \\
                        \hline
                        \texttt{learning\_rate} & 0 & 2 \\
                        \texttt{min\_split\_loss} & 0 & 2 \\
                        \texttt{max\_depth} & 0 & 256 \\
                        \texttt{min\_child\_weight} & 0 & 10 \\
                        \texttt{reg\_lambda} & 0 & 2 \\
                        \texttt{reg\_alpha} & 0 & 2 \\
                    \end{tabular}
                    \label{tab:methodology:ml-models:gradient-boosting-with-xgboost:hyperparameters}
                \end{table}
            
            The variable names here (e.g. \texttt{learning\_rate}) correspond with the variable names listed in the documentation of \ac{XGBoost}. Additionally, Optuna was used to select a standardization strategy, choosing between Z-score normalization (i.e. subtracting the mean and dividing by the standard deviation) or Min/Max scaling (i.e. scaling the data such that it has minimum 0 and maximum 1). To prevent test-set leakage, the chosen standardizer was fit only with the internal training set, i.e. the portion of the training set that was not held out as an internal validation set. Optuna performed 1000 trials to minimize the validation set \ac{RMSE}. We report the results of the final optimized model. 
        
        \subsubsection{AutoML with TPOT}
        \label{methodology:ml-models:automl-with-tpot}
            The \ac{AutoML} tool \ac{TPOT} was leveraged with a population size of 100 pipelines, with training proceeding for a total of 10 generations. A maximum evaluation time of 10 seconds per model was set. \ac{TPOT} pipelines were optimized using the default regression configuration. As \ac{TPOT} is an actively-maintained open-source repository, for the purposes of future replication we enumerate this configuration's set of allowable components in Table \ref{stab:tpot-model-components:model-selection}. The models listed in this table could be combined in any order any number of times. Models were selected such that their 10-fold cross-validated \ac{RMSE} was optimized. \ac{TPOT} also conducts its own internal optimization of model hyperparameters, thus we did not perform our own hyperparameter optimization of the \ac{TPOT} pipelines.
        
        \subsubsection{Neural Networks with Roost}
        \label{methodology:ml-models:nn-with-roost}
            The \ac{Roost} \ac{NN} architecture was leveraged using the ``example.py'' script provided with its source code. Models are trained for a total of 512 epochs with the default settings. In the case of \ac{Roost} models, the only feature provided is the composition of the system, given through the chemical formula.
        
        \subsubsection{Symbolic Regression with SISSO}
            % \todo{Should we mention here that we even tried contacting XenonPy on their (fairly active) Github issues page? Should also mail the corresponding author of the paper, so we can report that we exhausted all options of finding this data's provenance}
            
            % \TP{I wouldn't, just state what we are using and how we came to those unit sets}
            
            % \JD{Makes sense, probably won't be doing that then.}
            
        \label{methodology:ml-models:symbolic-regression-with-sisso}
            The first step of using SISSO is reducing the number of primary features down from a list of hundreds down to the tens. This is done due to the exponential computational cost of \ac{SISSO} with respect to the number of features and the number of rungs being considered. To perform this down selection we first generate a rung 1 feature space including all of the primary features and operators that are used in the SISSO calculation. We then check how often each of the primary features appear in the ten thousand generated features that are most correlated to the target property. Additionally, we add units to all of the pre-selected primary features to ensure all generated expressions are valid.
            % To do this we use two approaches to reduce the number of features: down-selecting with LASSO and prescreening the primary features by checking how often they are included in the most correlated rung 1 features. For the LASSO down-selection we take the 16 features found to be the most important by LASSO. For the SISSO prescreening we generate a rung 1 feature space including all of the primary features and operators that are used in the SISSO calculation, and then take the 30 features that appear the most often in the 10,000 features that are most correlated to the target property. Additionally we add units to all of the pre-selected primary features to ensure all generated expressions are valid.
            % For models trained using \ac{SISSO}, we first perform feature selection via \ac{LASSO}, down-selecting to a set of 16 features. This is done due to the exponential computational cost of \ac{SISSO} with respect to the number of features and the number of rungs being considered. In order to create more-interpretable features, \ac{SISSO} is capable of considering the units of the different input features when creating new compound features.
            
            % \TP{We may want to cut this section because this might be seen as an attack on XenonPy. I would phrase it neturally, i.e. For some cases where units were not reported we compared values to elements in known sources (e.g. the NIST WebBook \cite{linstromNISTChemistryWebBook2021} or the CRC Handbook \cite{rumbleCRCHandbookChemistry2021}) and determined what the units are.}
            % \JD{ Rephrased it to be a lot more neutral}
            
            In many cases, it was easy to infer what the abstract units are for the XenonPy descriptors. In a few cases where the units weren't as clear, we compared the reported elemental values of those units to those of known sources (e.g. the NIST WebBook \cite{linstromNISTChemistryWebBook2021} or the CRC Handbook  \cite{rumbleCRCHandbookChemistry2021}) in order to determine the units. Finally, although it was generally easy to determine where the source of a feature was, sometimes we were unable to determine a source. In these cases, we refer to the features as a "XenonPy" feature (for example,  ``\(r_{XenonPy}\)'').
            
            % \TP{I would rephrase to sound less of an attack on XenonPy. I would say: In some cases where the source of the initial features could not be verified, we renamed the features to be, ``\(r_{XenonPy}\)''.} 
            % \JD{Totally removed the section and collapsed into the above paragraph}
            
            % \TP{I used a SIS of 100 for my runs}
            % Per TB's instruction, we plan to hold off on including those runs in the first arXiv publication, and stick with what we have now to publish to arXiv faster. They should be included in the next iteration of the paper. -JD
            
            The optimal number of terms (up to 3) and rung (up to 2), i.e. the number of times operators are recursively applied to the feature space, are determined using a five-fold cross validation scheme. For all models we allow for an external bias term to be non-zero, and use a SIS selection size of 500. The resulting descriptors were then evaluated using the same external test set for each of the other methods. To take advantage of \ac{SISSO}'s ability to generate new composite descriptors and operate in large feature spaces, additional features were included in the \ac{SISSO} calculations. A full list of features used in the SISSO work can be found in the linked GitHub repository.
             
            % Models with up to 4 terms were considered, and the initial SIS selection chose 10 features. An intercept was not fixed. An internal validation set of 10\% of the data was held out. SISSO models up to Rung 2 are considered, with Rung 1 and Rung 2 being explored in separate runs.

\section{Results}
\label{sec:results}
    % ==================
    % Perovskite Volume
    % ==================
    \subsection{Perovskite Volume Prediction}
    \label{results:perovskite-volume-prediction}
\begin{figure}[H]
            \centering
            \includegraphics[width=\textwidth,height=\textheight,keepaspectratio]{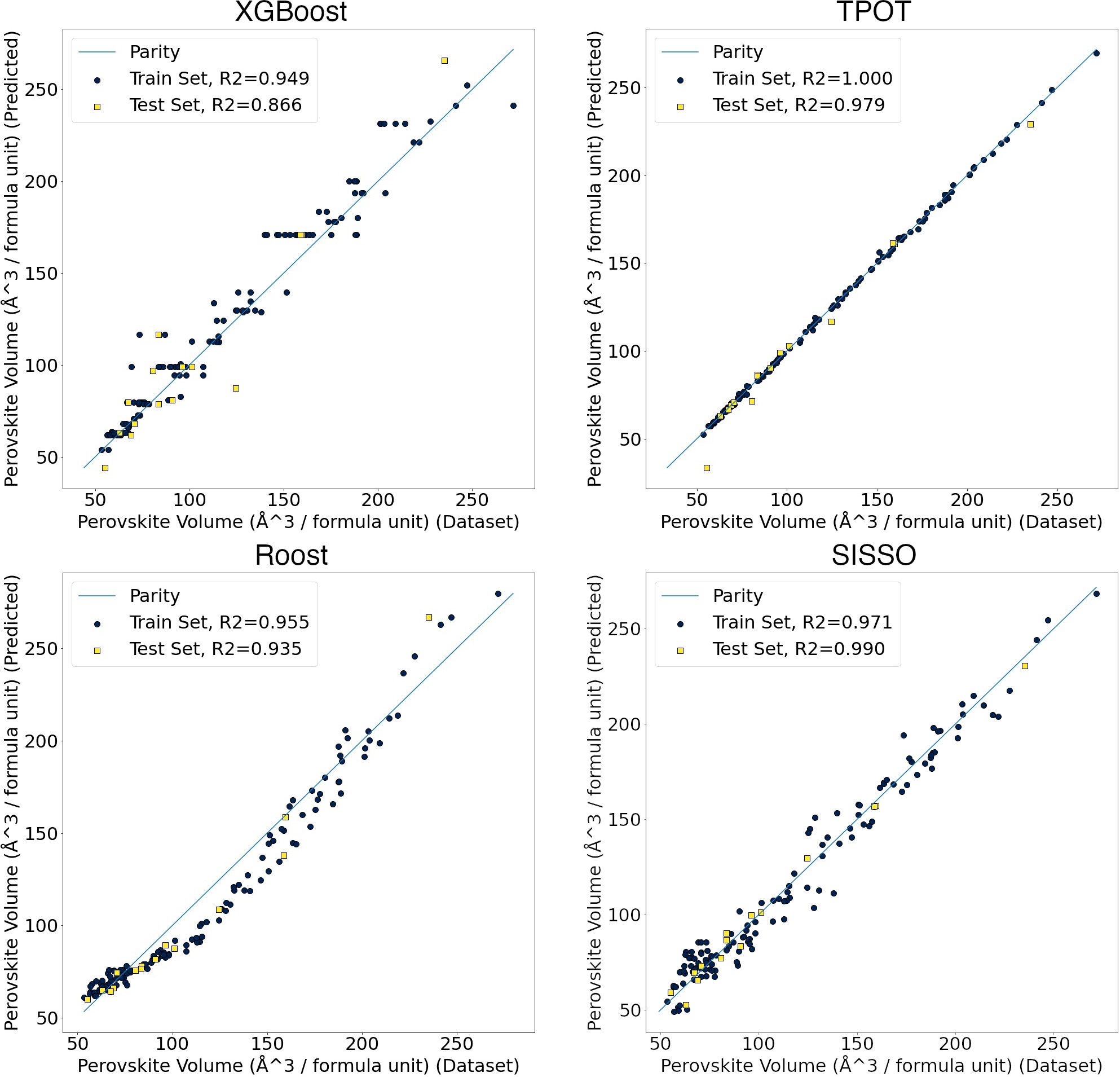}
            \caption{Parity plots for the \ac{XGBoost}, \ac{TPOT}, \ac{Roost}, and \ac{SISSO} models on the perovskite unit cell volume problem. Included are the training and testing sets. A diagonal line indicating parity is drawn as a guide to the eye. 
            % The SISSO model we report is the rung 1, 4-term model. Regression statistics for the models shown on this plot can be found in Table \ref{tab:results:perovskite-tpot-sisso-comparison}.
            }
            \label{fig:results:perovskite-tpot-sisso-comparison}
        \end{figure}

        \ac{XGBoost}, \ac{TPOT}, and \ac{SISSO} were applied to investigate the volume of perovskites as a function of the compositional features described in section \ref{methodology:feature-engineering:compositional-descriptors}. Additionally, we trained a \ac{Roost} model on the chemical formula of the perovskites to predict the volume. The train/test split resulted in a total of 129 entries in the training set, and 15 in the test set. We find generally good performance on the perovskite volume problem across all 4 models, although the \ac{TPOT} and \ac{SISSO} model display the best performance by all metrics investigated (see Table \ref{tab:results:perovskite-tpot-sisso-comparison}), including respective test-set R\textsuperscript{2} of 0.979 and 0.990. 
        The \ac{Roost} model also performs well with a test-set R\textsuperscript{2} of 0.935, but it also has a non-normal error, as can be seen in Figure~\ref{fig:results:perovskite-tpot-sisso-comparison}. Finally, we find that while \ac{XGBoost} is the worst performing method, it still has a relatively good test-set R\textsuperscript{2} of 0.866.
        % The next best-performer was the \ac{Roost} model, with a superior \ac{RMSE}, Max Error, and R\textsuperscript{2} on the training and set sets compared to the other models. Finally, we find that the \ac{XGBoost} and \ac{SISSO} models had similar performance, but still with good test-set R\textsuperscript{2} of 0.866 and 0.823 respectively. 
        % \TP{Are we not updating all of the SISSO results?}
        % Not currently; based on the direction of TB we are keeping what we have to facilitate a faster time to arXiv.
        \begin{table}[H]
            \centering
            \rowcolors{2}{gray!25}{white}
            \caption{Performance metrics for the \ac{XGBoost}, \ac{TPOT}, \ac{Roost}, and \ac{SISSO} models on the perovskite volume prediction problem. 
            % The rung-1, 4-term \ac{SISSO} model is reported. 
            The parity plots for these models are depicted in \ref{fig:results:perovskite-tpot-sisso-comparison}.}
            \begin{tabular}{rccccc}
                \rowcolor{gray!50}
                Error Metric & Partition & \acs{XGBoost} & \acs{TPOT} & \acs{Roost} & \ac{SISSO}\\
                \hline
                \acs{MAE}            & Train & 8.15   & 0.860  & 9.11   & 7.27 \\
                \acs{RMSE}           & Train & 11.88  & 1.14  & 11.21  & 9.08 \\
                Max Error            & Train & 43.29  & 4.96  & 22.94  & 26.74 \\
                R\textsuperscript{2} & Train & 0.950  & 1.000 & 0.955  & 0.971 \\
                \hline
                \acs{MAE}            & Test & 12.89   & 4.02  & 8.83   & 4.05 \\
                \acs{RMSE}           & Test & 17.17   & 6.78  & 12.00  & 4.71 \\
                Max Error            & Test & 37.10   & 21.75 & 31.69  & 10.27 \\
                R\textsuperscript{2} & Test & 0.866   & 0.979 & 0.935  & 0.990 \\
            \end{tabular}
            \label{tab:results:perovskite-tpot-sisso-comparison}
        \end{table}
        
        The performance of all 4 models is summarized in Figure \ref{fig:results:perovskite-tpot-sisso-comparison}. Visually, we find a very tight fit by the \ac{TPOT} model in both the training and test sets, with good correlation from the \ac{XGBoost} and \ac{SISSO} models. We also find a systematic under-prediction of perovskite volumes in the roost model in both the training and test set, with the under-prediction beginning at approximately 75 Å\textsuperscript{3} / formula unit, achieving a maximum deviation at approximately 130 Å\textsuperscript{3} / formula unit, and returning to parity at approximately 200Å\textsuperscript{3} / formula unit.

        The good performance of the \ac{TPOT} model results from a generated pipeline with four stages. The first two stages are \texttt{MinMaxScaler} units; although the second one is redundant, as the data is already scaled to be between 0 and 1 by the first scaler. The third stage is a \texttt{OneHotEncoder} unit, which leverages one-hot encoding for categorical features (the \ac{TPOT} implementation defines a categorical feature as one with fewer than 10 unique values). Finally, the perovskite volume is predicted using support vector regression.
        
        In the case of the \ac{XGBoost} model, we can extract feature importances. Although various different feature importance metrics can be derived from \ac{XGBoost}, in this case we use the ``gain'' metric, which describes how the model's loss function improves when a feature is chosen for a split while constructing the trees. A large number of features (290) were input into this model, so we display only the 10 most-important features identified by \ac{XGBoost} in Supporting Information Figure \ref{sfig:xgboost-feature-importance:perovskite-xgboost-importances}. Here, we find that the average Rahm atomic radii \cite{rahmAtomicIonicRadii2016, rahmCorrigendumAtomicIonic2017} have the highest importance score, followed by the average Van der Waals radius used by the \ac{UFF} \cite{rappeUFFFullPeriodic1992}. The remaining 288 features fall off as a long tail of low importance scores, indicating that they did little to improve the model's performance in predicting the perovskite volume.
        
        For \ac{SISSO}, we used reduced the feature space as outlined in Section \ref{methodology:ml-models:symbolic-regression-with-sisso},
        with the pre-screened features listed in the Supporting Information %Table \ref{stab:sisso-feature-selection:perovskite-sisso-prevelances},
        along with the assumption we made about the units of the descriptor when fed into \ac{SISSO}. Generally, we find that the main descriptors selected by the procedure are related to volume and atomic radius. Some other descriptors with less interpretability are found, such as the C6 dispersion coefficients, polarizability, melting points, and \ac{HHI}\cite{gaultoisDataDrivenReviewThermoelectric2013} production and reserve values. Although typically used to help indicate the size of a company within a particular sector of the economy, the XenonPy definition of \ac{HHI} appears to come from the work of Gaultois et al \cite{gaultoisDataDrivenReviewThermoelectric2013}. In the referenced work, the \ac{HHI} production value refers to the geographic distribution of elemental production (e.g. answering the question of concentrated the industry that produces pure Fluorine is), and \ac{HHI} reserve value describes the geographic distribution of known deposits of these materials (e.g. whether they are spread out over a wide area, or concentrated in a small area).
        
        % For \ac{SISSO}, we used \ac{LASSO}-based feature selection to identify 16 features for use in the model. These are reported in Supporting Information Table \ref{stab:perovskite-sisso-importances}, along with the assumption we made about the units of the descriptor when fed into \ac{SISSO}. Generally, we find that the main descriptors selected by \ac{LASSO} are related to volume and atomic radius. Some other descriptors with less interpretability are found, such as the C6 dispersion coefficients, sound velocity, melting points, and \ac{HHI}\cite{gaultoisDataDrivenReviewThermoelectric2013} production and reserve values. Although typically used to help indicate the size of a company within a particular sector of the economy, the XenonPy definition of \ac{HHI} appears to come from the work of Gaultois et al \cite{gaultoisDataDrivenReviewThermoelectric2013}. In the referenced work, the \ac{HHI} production value refers to the geographic distribution of elemental production (e.g. answering the question of concentrated the industry that produces pure Fluorine is), and \ac{HHI} reserve value describes the geographic distribution of known deposits of these materials (e.g. whether they are spread out over a wide area, or concentrated in a small area).

        We report the best descriptor found in Equation \ref{eqn:results:perovskite-sisso}. In this equation, the variables \(c_0, a_0, a_1\) are the regression coefficients determined by \ac{SISSO}. 
        \begin{equation}
            \begin{multlined}
                V_{Perovskite} \approx c_0 + a_0 \cdot \frac{Z^{ave}}{C^{ave}\cdot \left(r^{ave}_{Slater} - r^{ave}_{pyykko, triple}\right)}+ a_1 \cdot \left(V^{ave}_{gs}-V^{min}_{gs}\right) \cdot \frac{r^{ave}_{pyykko, triple}}{r^{ave}_{pyykko}}
            \end{multlined}
            \label{eqn:results:perovskite-sisso}
        \end{equation}
        where $c_0 = -10.547$, $a_0 =4.556$, $c_1 =3.050$, $Z^{ave}$ is the average atomic number, $C^{ave}$ is the average mass specific heat capacity of the elemental solid, $r^{ave}_{Slater}$ is the average atomic covalent radius predicted by Slater, $r^{ave}_{pyykko, triple}$ is the average triple bond covalent radius predicted by Pyyko, $r^{ave}_{pyykko}$ is the average single bond covalent radius predicted by Pyyko, and $V^{ave}_{gs}$ and $V^{min}_{gs}$ are the average and minimum ground state volume per atom as calculated by DFT.
        Unsurprisingly the ground state atomic volumes and covalent radii play an important role in determining the final volume of the perovskite structures. Interestingly, both the atomic number and specific heat capacity of the material appear in the final descriptor; however, these likely only act as minor corrections to the final results.
        
        % We find three descriptors are being chosen by \ac{SISSO}: \(T_{melt}^{var}\) refers to the variance of the melting point, \(T_{boil}^{ave}\) refers to the average boiling point, and \(r_{Rahm}^{ave}\) refers to the average atomic radius of Rahm, Hoffmann, and Ashcroft \cite{rahmAtomicIonicRadii2016,rahmCorrigendumAtomicIonic2017}. The value of the intercept \(c_0\) is \(-48.92\), and the \(a_0-a_3\) terms are \(4.16*10^{-4}\), \(-1.91*10^{-6}\), \(-9.31*10^{-2}\), and \(2.17*10^{-5}\) respectively. Of interest is that one of the descriptors is the cube of the atomic radius. In addition to using the atomic radii, the \ac{SISSO} model seems to heavily rely on the melting and boiling point of the systems.
        
    % =======
    % Bandgaps
    % ========
    \subsection{2D Material Bandgaps}
    \label{results:2d-material-bandgaps}
        The bandgap predictions leveraged a data filtering strategy (described in Section \ref{methodology:data-filtering}. As a result of our data filtering approach, the 6,351 entries in the dataset were reduced to 1,412 entries. The train/test split divided the data into a training set of 1,270 rows, and a test set containing 142 entries. The performance metrics of the \ac{XGBoost}, \ac{TPOT}, \ac{Roost}, and \ac{SISSO} models of 2D Material Bandgap can be found in Table \ref{tab:results:bandgap-2d-tpot-sisso-roost-xgboost}. Performance is generally worse on this problem when compared to the perovskite volume predictions. As a result, in addition to the compositional features of XenonPy (Section \ref{methodology:feature-engineering:compositional-descriptors}) we also used several structural features (section \ref{methodology:feature-engineering:structural-descriptors:matminer-descriptors}). We also leveraged the bulk bandgap of the parent-3D material for each of the 2D materials, as we observed the performance of the \ac{TPOT}, \ac{SISSO} and \ac{XGBoost} models increased when this value was included.
        
        \begin{table}[H]
            \centering
            \rowcolors{2}{gray!25}{white}
            \caption{Performance metrics for the \ac{XGBoost}, \ac{TPOT}, \ac{Roost}, and \ac{SISSO} models on the 2D material bandgap problem. The rung-2, 4-term \ac{SISSO} model is reported. The parity plots for these models are depicted in \ref{fig:results:bandgap-2d-tpot-sisso-roost-xgboost}.}
            \begin{tabular}{rccccc}
                \rowcolor{gray!50}
                Error Metric & Partition & \acs{XGBoost} & \acs{TPOT} & \acs{Roost} & \ac{SISSO}\\
                \hline
                \acs{MAE}            & Train & 0.12  & 0.27  & 0.11  & 0.29   \\
                \acs{RMSE}           & Train & 0.25  & 0.41  & 0.27  & 0.47   \\
                Max Error            & Train & 2.75  & 3.66  & 2.83  & 4.35   \\
                R\textsuperscript{2} & Train & 0.973 & 0.928 & 0.968 & 0.907  \\
                \hline
                \acs{MAE}            & Test & 0.31   & 0.29   & 0.65  & 0.29  \\
                \acs{RMSE}           & Test & 0.50   & 0.48   & 1.07  & 0.47  \\
                Max Error            & Test & 2.04   & 2.75   & 4.80  & 3.27  \\
                R\textsuperscript{2} & Test & 0.892  & 0.900  & 0.507 & 0.919 \\
            \end{tabular}
            \label{tab:results:bandgap-2d-tpot-sisso-roost-xgboost}
        \end{table}
        
        Although test-set model performance was worse compared to the perovskite problem, the \ac{TPOT} and \ac{SISSO} models retained their status as having the best performance metrics for the test-set R\textsuperscript{2}, \ac{MAE}, and \ac{RMSE}. The \ac{XGBoost} model additionally performed well, with the best performance in terms of maximum error, and nearly as good performance on the other metrics when compared to \ac{TPOT} and \ac{SISSO}. We find the \ac{Roost} model overfit the data to some extent on the data, as the test-set error metrics are considerably worse than their training-set counterparts.% (based on the gap in performance between the training and testing sets). In this case, we find the \ac{SISSO} model to have had the worst performance of the systems.
        
\begin{figure}[H]
            \centering
            \includegraphics[width=\textwidth,height=\textheight,keepaspectratio]{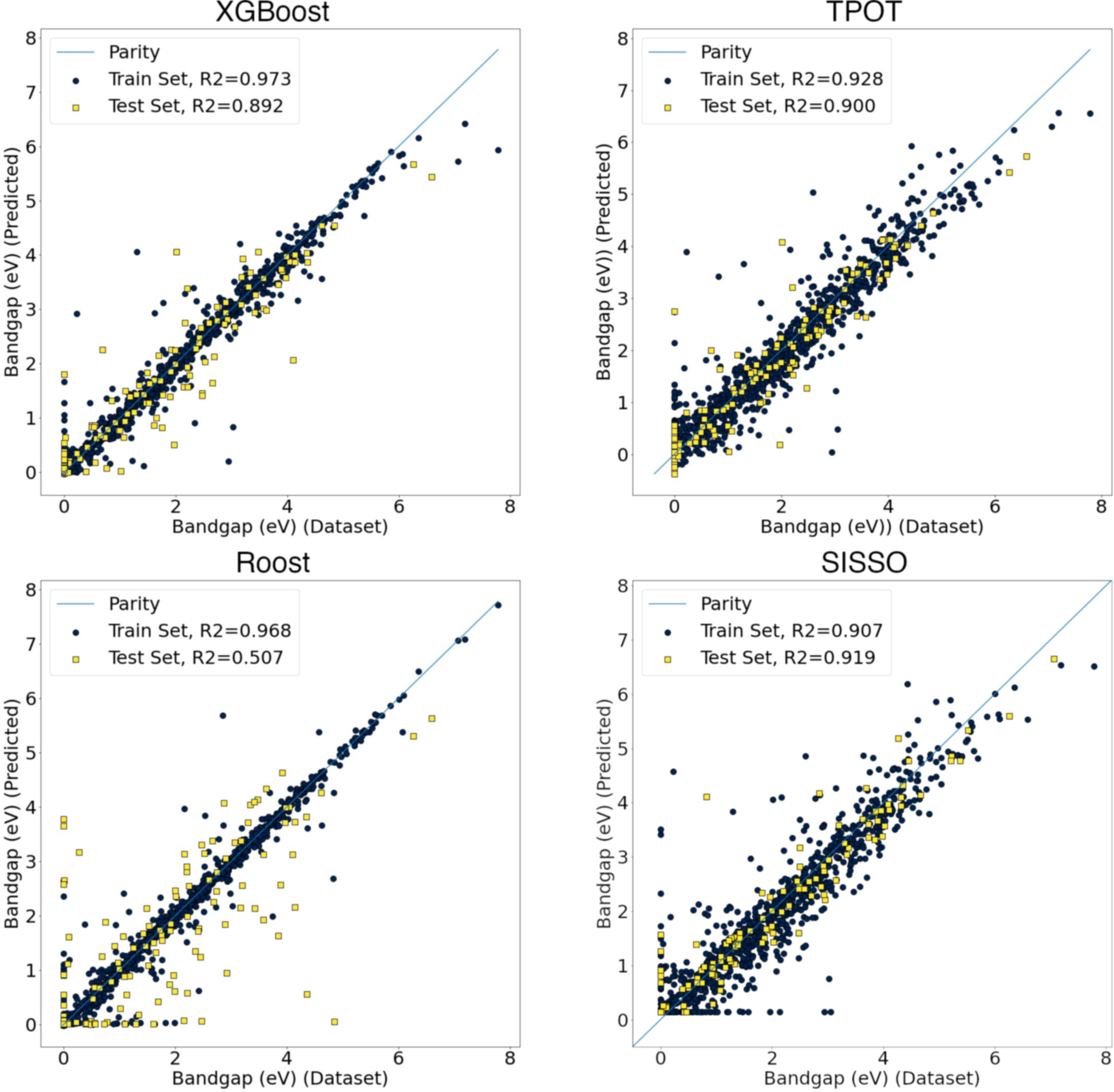}
            \caption{Parity plots for the \ac{XGBoost}, \ac{TPOT}, \ac{Roost}, and \ac{SISSO} models on the 2D  material bandgap problem. Included are the training and testing sets. A diagonal line indicating parity is drawn as a guide to the eye. %The SISSO model we report is the rung 2, 4-term model. 
            Regression statistics for the models shown on this plot can be found in Table \ref{tab:results:bandgap-2d-tpot-sisso-roost-xgboost}.}
            \label{fig:results:bandgap-2d-tpot-sisso-roost-xgboost}
        \end{figure}

        A parity plot summarizing these results can be found in Figure \ref{fig:results:bandgap-2d-tpot-sisso-roost-xgboost}. In all cases, we can see a spike of misprediction for systems with a \ac{DFT} bandgap of 0. We note here that a large portion of these entries had \ac{DFT} bandgaps of 0: of the 382 of the 1,412 entries in the dataset, a total of 27\% of all training data.
        
        The pipeline generated by \ac{TPOT} is less complex than that of the perovskite volume problem. The first stage of the pipeline is a \texttt{MinMaxScaler} unit, scaling each feature such that it is between 0 and 1. The second stage is then an \texttt{ElasticNetCV} unit, which uses 5-fold cross-validation to optimize the alpha and L1/L2 ratio of the Elastic Net model. The converged alpha value was \(1.011*10^{-3}\), and the converged L1/L2 ratio was 0.95, which strongly leans towards the L1 (\ac{LASSO}) regularization penalty.
        
        We can also extract feature importances from the \ac{XGBoost} model, and we report the 10 highest-ranked features in Supporting Information Figure \ref{sfig:xgboost-feature-importance:bandgap-xgboost-importances}. In contrast with the perovskite results, the features are generally ranked similarly; although there is a clear ranking, we do not see 1-2 features dominating followed by a long tail of unimportant features. The selected features are also less interpretable; we find the minimum Mendeleev number as the most important feature, an alternative numbering of the periodic table in which numbers are ordered by their group rather than period. Specifically, we use the variant proposed by Villars et al \cite{villarsDatadrivenAtomicEnvironment2004}. For example, in the referenced system of numbering, Li, Na, K, and Rb are elements 1, 2, 3, and 4. This numbering is generally consistent, with the exception is H, which is assigned a value of 92 to place it above F, which has value 93. By taking the minimum value of the composition, this metric can be intuited as identifying the element in the earliest group and earliest period of the periodic table. Other descriptors we find are the average atomic weight, the variance of the number of unfilled s-orbitals in the system, and \ac{DFT}-calculated ground-state atomic volumes.
        
        The results of the prescreening procedure for \ac{SISSO} are presented in the Supporting Information Table~\ref{stab:bandgap-sisso-importances}.
        These features are similar to the previous results, with the addition of the parent-3D material bandgap and electronegativity information playing an important role as well.
        % The \ac{LASSO} feature selection is presented in Supporting Information Table \ref{stab:bandgap-sisso-importances}. We find generally similar features to those used in the perovskite volume prediction problem (Supporting Information Table \ref{stab:perovskite-sisso-importances}) --- atomic radii, volumes, melting points, etc. Additionally, we find that the polarizability and thermal conductivity are selected as potential features as well.
        
        The selected \ac{SISSO} model is
        \begin{equation}
            E^{2D}_{Bandgap} \approx c_0 + a_0 \cdot \frac{r^{min}_{vdw} \cdot E_{Bandgap}^{3D, parent}}{\sqrt[3]{Period^{min}}}
            \label{eqn:results:bandgap-sisso}
        \end{equation}
        where $c_0=0.143$, $a_0=8.054\times 10^{-3}$, $r^{min}_{vdw}$ is the minimum Van der Waals radius of the atoms in the material, $E_{Bandgap}^{3D, parent}$ is the bandgap of the 3D-parent material, and $Period^{min}$ is the minimum period of the elements in the material.
        This descriptor represents a simple rescaling of the bandgap of the 3D-parent material, which makes sense as both bandgaps are highly correlated to each other.
        Furthermore this results is consistent with the \ac{TPOT} model which is also primarily controlled by the bandgap of the 3D-parent material.

    % ==================
    % Exfoliation Energy
    % ==================
    \subsection{2D Material Exfoliation Energy}
    \label{results:2d-material-exfoliation-energy}
        In the case of the 2D material exfoliation energy problem, the training and test-set statistics for the \ac{XGBoost}, \ac{TPOT}, \ac{Roost}, and \ac{SISSO} models can be found in Table \ref{tab:results:exfoliation-2d-tpot-sisso-roost-xgboost}. In this case, our feature selection methodology down-selected the 6,351 rows of our dataset into 3,388 rows. The train/test split further divided this into a training-set of 3,049 entries, and a test set of 339 entries. Generally, we see the worst performance of the models in this problem, compared to the perovskite volume and 2D material bandgap problems.
        
        \begin{table}[H]
            \centering
            \rowcolors{2}{gray!25}{white}
            \caption{Performance metrics for the \ac{XGBoost}, \ac{TPOT}, \ac{Roost}, and \ac{SISSO} models on the 2D material exfoliation energy problem. 
            % The rung-2, 2-term \ac{SISSO} model is reported. 
            The parity plots for these models are depicted in \ref{fig:results:exfoliation-2d-tpot-sisso-roost-xgboost}.}
            \begin{tabular}{rccccc}
                \rowcolor{gray!50}
                Error Metric & Partition & \acs{XGBoost} & \acs{TPOT} & \acs{Roost} & \ac{SISSO}\\
                \hline
                \acs{MAE}            & Train & 0.20  & 0.14  & 0.06  & 0.26  \\
                \acs{RMSE}           & Train & 0.35  & 0.31  & 0.24  & 0.45  \\
                Max Error            & Train & 7.11  & 8.34  & 9.63  & 9.58  \\
                R\textsuperscript{2} & Train & 0.624 & 0.702 & 0.827 & 0.365 \\
                \hline
                \acs{MAE}            & Test & 0.23   & 0.21  & 0.19  & 0.27  \\
                \acs{RMSE}           & Test & 0.35   & 0.33  & 0.34  & 0.42  \\
                Max Error            & Test & 1.64   & 1.85  & 1.96  & 2.48  \\
                R\textsuperscript{2} & Test & 0.476  & 0.543 & 0.499 & 0.2412 \\
            \end{tabular}
            \label{tab:results:exfoliation-2d-tpot-sisso-roost-xgboost}
        \end{table}
        
\begin{figure}[H]
            \centering
            \includegraphics[width=\textwidth,height=\textheight,keepaspectratio]{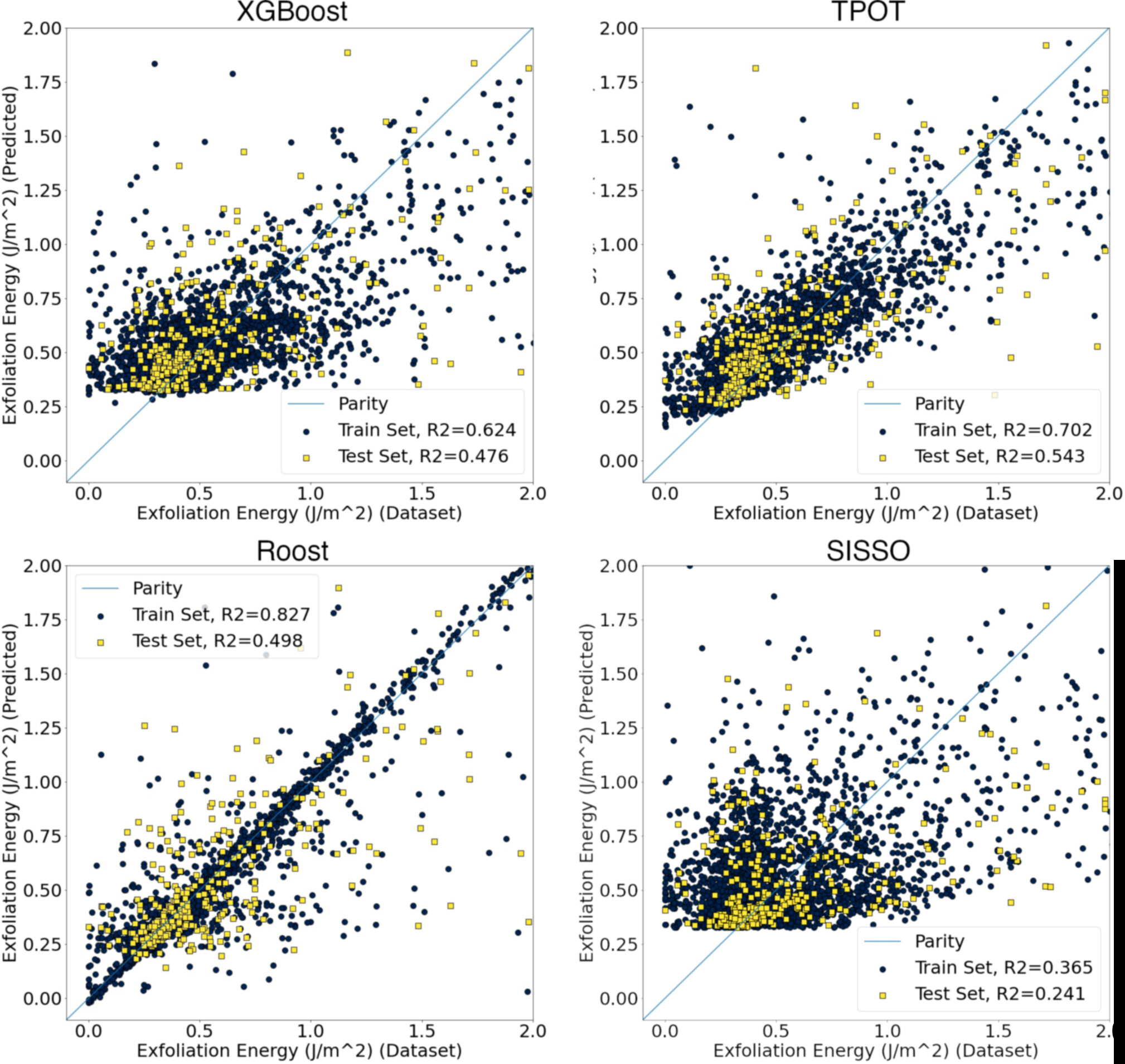}
            \caption{Parity plots for the \ac{XGBoost}, \ac{TPOT}, \ac{Roost}, and \ac{SISSO} models on the 2D  material exfoliation energy problem. Included are the training and testing sets. A diagonal line indicating parity is drawn as a guide to the eye. 
            % The SISSO model we report is the rung 2, 2-term model. 
            Regression statistics for the models shown on this plot can be found in Table \ref{tab:results:exfoliation-2d-tpot-sisso-roost-xgboost}. To facilitate comparison at energy ranges that are more experimentally-relevant, we have zoomed in the plot to study energies no higher than 2 eV. The full data range is plot in Supporting Information Figure \ref{sfig:exfoliation-energy-comparison:exfoliation-2d-tpot-sisso-roost-xgboost}.}
            \label{fig:results:exfoliation-2d-tpot-sisso-roost-xgboost}
        \end{figure}

        A set of parity plots for all four models is presented in \ref{fig:results:bandgap-2d-tpot-sisso-roost-xgboost}. To facilitate easier comparison at experimentally-relevant energy ranges, we have zoomed the plot in such that the highest exfoliation energy is 2 eV. Plots showing the entire energy range explored can be found in Supporting Information Figure \ref{sfig:exfoliation-energy-comparison:exfoliation-2d-tpot-sisso-roost-xgboost} Here, we find that all models perform generally poorly, with the largest errors occurring at higher exfoliation energies in the case of \ac{XGBoost}, \ac{TPOT}, and \ac{SISSO}. (see Figure \ref{fig:results:exfoliation-2d-tpot-sisso-roost-xgboost}). The best test-set R\textsuperscript{2} and \ac{RMSE} this time is only \ac{TPOT}, although they are still relatively poor, with a test-set R\textsuperscript{2} of only 0.543. \ac{Roost} displays the best test-set \ac{MAE}, although the model seems to have overfit, as it displays drastically better performance on the training set than it does on the test set. The \ac{XGBoost} model performs slightly worse than either \ac{TPOT} or \ac{Roost}, and the \ac{SISSO} approach did not perform well for this problem
    
        The \ac{TPOT} algorithm again results in a relatively simple model pipeline. The first stage is a \texttt{SelectFwe} unit, which down-selects the features according to the \ac{FWE} \cite{hastieElementsStatisticalLearning2009}. An alpha value of \(0.047\) is selected for this purpose, with results in a down-selection of the 299 input features into 125 features. This is then fed into an \texttt{ExtraTreesRegressor} unit, which is an implementation of the Extremely Randomized Trees method proposed by Geurts, Ernst, and Wehenkel\cite{geurtsExtremelyRandomizedTrees2006}.
        
        We again extract features from the \ac{XGBoost} model (Supporting Information Figure \ref{sfig:xgboost-feature-importance:exfoliation-xgboost-importances}), and find the Mendeleev Number again appears as an important feature, albeit as the maximum instead of the minimum. Additionally, we see descriptors related to bond strengths in the corresponding elemental systems: average melting points, and average heats of evaporation.
        
        The list of preselected features can be found in the Supporting Information Table \ref{stab:exfoliation-sisso-importances}. Overall, we see that this set of features is the least similar out of all three problems with the bulk modulus, thermal conductivity, and decomposition energy being the most prevalent. Additionally, the bandgap of the material is also selected, suggesting the possible hierarchical learning. For this calculation additional features such as the decomposition energy, ratio between the perimeter and area of the surface, and the electronic bandgap of the material are also included.
        
        The best SISSO model found for this problem is
        \begin{equation}
            \begin{multlined}
                E_{Exfoliation} \approx c_0 + a_0 \cdot \frac{P}{A} \cdot q_{evaporation}^{min} \cdot \sqrt{\sum_i\kappa_i} + a_1 \cdot \frac{E_{decomposition}}{V_{ICSD}^{ave} \cdot {r_{vdw,uff}^{min}}^3}
            \end{multlined}
            \label{eqn:results:exfoliation-sisso-rung2-2term}
        \end{equation}
        where $c_0=0.327$, $a_0=1.24\times 10^{-4}$, $a_1=9.26\times 10^8$, $\frac{P}{A}$ is the perimeter to area ratio of the surface, $q_{evaporation}^{min}$ is the minimum atomic evaporation heat of each element in the material, $E_{decomposition}$ is the decomposition energy, $\kappa_i$ is the thermal conductivity of the elements at 298 K, $V_{ICSD}^{ave}$ is the average atomic volume in the ICSD database, and $r_{vdw,uff}^{min}$ is the minimum atomic Van der Waals radius from UFF for each element in the material.

\section{Discussion}
\label{sec:discussion}
    We have developed a series of models which are capable of generating predictions for (1) the volume per formula unit of a series of ABX\textsubscript{3} perovskites, (2) the \ac{DFT}-calculated bandgap of several 2D materials, and (3) several 2D material exfoliation energies. These problems encompass a variety of outcomes that one may find when training models of predictive properties.
    
    \subsection{Perovskite Volume per Formula Unit}
    \label{discussion:perovskite-volume-per-formula-unit}
        In the case of the volume per formula unit of ABX\textsubscript{3} perovskites, we observe all four model types perform well. Overall, we find that the volume per formula unit for ABX\textsubscript{3} perovskites can be predicted using only compositional descriptors (i.e. with no structural descriptors). The likely reason all four models perform well despite having no structural information is the general similarity in crystal structure between these systems --- they are all perovskites, and therefore all possess very similar crystal structures. Supporting this is that the \ac{Roost} model, which only leverages the chemical formula as an input, and which we did not optimize the hyperparameters or architecture for, performed just as well on this problem --- albeit with some systematic deviation from parity at intermediate volumes. Although interpretability is reduced by virtue of being a neural network, we can still achieve an important insight from this model --- just by knowing the chemical formula of the system, we can achieve accurate predictions of perovskite volumes, which further justifies our use of compositional descriptors (see Section \ref{methodology:feature-engineering:compositional-descriptors}) on this problem as we move to the \ac{SISSO}, \ac{XGBoost}, and \ac{TPOT} models. Additionally, we note this performance was achieved with a training set containing only 129 entries in the dataset - compared to the original \ac{Roost} paper\cite{goodallPredictingMaterialsProperties2020} that leveraged approximately 275,000 entries from the \ac{OQMD} datset\cite{jhaElemNetDeepLearning2018}.
        
        Like the \ac{Roost} model, we have difficulty in interpreting the pipeline generated by \ac{TPOT}. The \ac{TPOT} model delivers the best performance --- which is clearly visible from the parity plot in Figure \ref{fig:results:perovskite-tpot-sisso-comparison}. This performance came at a price, however, and the rather complex pipeline containing scaling, one-hot encoding, and linear support vector regression does not lend itself well to interpretation.
        
        Entering into the realm of interpretability, although the \ac{XGBoost} model does not produce a direct formula for perovskite volumes, we can still gain some insight using it. It is still, however, relatively accurate --- and allows us access to a feature importance metric (see Supporting Information Figure \ref{sfig:xgboost-feature-importance:bandgap-xgboost-importances}). In this case, we see the five most-important features are the average Rahm\cite{rahmAtomicIonicRadii2016, rahmCorrigendumAtomicIonic2017} atomic radius, average \ac{UFF}\cite{rappeUFFFullPeriodic1992} atomic radius, sum of elemental velocities of sound in the material, average Ghosh\cite{ghoshNewScaleElectronegativity2005} electronegativity, and the sum of the Pyykko\cite{pyykkoTripleBondCovalentRadii2005} triple-bond covalent radii. Overall, we see a strong reliance on descriptors of atomic radius --- which as we noted in the \ac{TPOT} discussion makes intuitive sense.
        
        Finally, the \ac{SISSO} model (Equation \ref{eqn:results:perovskite-sisso}) offers the most direct interpretation, as it is simply an equation. Immediately, we can see that ground state DFT atomic volumes are important. This result is highly intuitive and is not surprising when we consider that i) we are predicting volume, therefore using an average atomic volume makes sense and ii) the \ac{TPOT} and \ac{XGBoost} models extensively leveraged atomic radius descriptors that are also related to the volume of the atoms.
        
         The overall good performance of SISSO for this application is promising, as it is one of the most accurate models, while being by far the most interpretable. This represents a key advantage to symbolic regression, as if you can find an accurate model, then it will be easy to understand and analyze the results. Moreover, we note that are not alone in the literature when it comes to leveraging \ac{SISSO} to generate models of perovskite properties --- the last several years have seen success in the creation of models of perovskite properties with this tool. The work of Xie et al. \cite{xieMachineLearningOctahedral2020} achieved good success in predicting the octahedral tilt in ABO\textsubscript{3} perovskites, the work of Bartel et al. \cite{bartelNewToleranceFactor2019} resulted in the creation of a new tolerance factor for ABX\textsubscript{3} perovskite formation, and Ihalage and Hao \cite{ihalageAnalogicalDiscoveryDisordered2021} leveraged descriptors generated by SISSO to predict the formation of quaternary perovskites with formula \((\text{A}_{1-x}\text{A}^\prime_x)\text{BO}_3\) and \(\text{A}(\text{B}_{1-x}\text{B}^\prime_x)\text{O}_3\).

    \subsection{2D Material Bandgap}
    \label{discussion:2d-material-bandgap}
        The 2D material bandgap models did not achieve the same performance as for the perovskite systems (see Figure \ref{fig:results:bandgap-2d-tpot-sisso-roost-xgboost}). Even in the case of \ac{TPOT} and \ac{SISSO}, which still had the best test-set performance by most metrics, there were a few outliers. Specifically, we find that the test-set \ac{MAE} for the models ranged between 0.29 eV (\ac{TPOT} and \ac{SISSO}) and 0.65 eV (\ac{Roost}) relative to the PBE \ac{DFT} calculations reported by the \ac{2DMatPedia}. Putting this number in perspective, we note the recent work of Tran et al\cite{tranBandgapTwodimensionalMaterials2021}, which benchmarked the bandgap predictions of several popular \ac{DFT} functions for many of the systems in the \ac{C2DB}; the work identified that the PBE functional exhibited a \ac{MAE} of 1.50 eV relative to the G\textsubscript{0}W\textsubscript{0} method. Other investigators have studied the prediction of 2D material bandgaps: Rajan et al.\cite{rajanMachineLearningAssistedAccurateBand2018} also achieved a test-set \ac{MAE} of 0.11 eV on a dataset of 23,870 MXene systems (which, as far as we are aware, has not been made publicly available) using a Gaussian Process regression approach, with \ac{DFT}-calculated properties including the average M-X bond length, volume per atom, MXene phase, fand heat of formation, and compositional properties including the mean Van der Waals radius, standard deviation of periodic table group number, standard deviation of the ionization energy, and standard deviation of the meting temperature. Zhang et al.\cite{zhangBandgapPredictionTwodimensional2021} improved on this error slightly, achieving a test-set \ac{MAE} of 0.10 eV on the \ac{C2DB} dataset (around 4000 entries)\cite{gjerdingRecentProgressComputational2021} with both Support-Vector Regression and Random Forest approaches, albeit using descriptors such as the fermi-energy density of states and total energy of the system (requiring further \ac{DFT} work for additional prediction). In contrast to both approaches, which used \ac{DFT}-calculated values that would need to be obtained for new systems to be predicted, the only \ac{DFT}-calculated value we leverage in our feature set is a bulk bandgap tabulated on the Materials Project\cite{jainCommentaryMaterialsProject2013}. Thus, although our \ac{TPOT} model had a slightly higher \ac{MAE} of 0.29 eV, we note that this would not require further \ac{DFT} work to generate new predictions. As 2D systems are still relatively new, we note that much more work has been performed in the 3D materials space, particularly in the leveraging of neural networks to predict bandgaps. The recent \ac{ALIGNN}\cite{choudharyAtomisticLineGraph2021} reported a test-set \ac{MAE} of 0.218 eV for the prediction of bulk materials hosted by Materials Project\cite{jainCommentaryMaterialsProject2013} (which as of October 2021 has over 144,000 inorganic systems). The \ac{MEGNet} architecture\cite{chenGraphNetworksUniversal2019} achieved a test-set \ac{MAE} of 0.32 eV on the bulk systems of the Materials Project. Although these neural network models are on 3D systems, we note that they do not leverage \ac{DFT} properties (which we re-iterate would cause any resulting model to require a \ac{DFT} calculation for future prediction) and had access to much larger datasets than the training set we obtained after filtering the \ac{2DMatPedia} entries (see Section \ref{methodology:data-filtering}). Overall, although the systems we investigate are not 3D bulk systems, we believe this puts the \ac{TPOT} \ac{MAE} of 0.29 eV for the bandgap of 2D systems in perspective.
        
        In all 4 models we trained, many of the incorrect predictions occur where the \ac{DFT} bandgap is 0 eV (which represented 27\% of the training set values). Because of this, we tried simplifying the bandgap problem, by training an \ac{XGBoost} model to predict whether the system was a metal (see Supporting Information section \ref{supporting-information:metal-classification-with-xgboost}), and showed that we could achieve good results --- although we incorporated a purely structural fingerprint, the Sine Matrix Eigenspectrum (see Supporting Information Section \ref{si:metal-classification-with-xgboost:structural-fingerprints}). As this descriptor resulted in some rather large vectors (of length 40, the maximum number of atoms in any system) with little direct physical intuition, we do not directly include it for the purposes of this section. Instead, what we can derive from it is the knowledge that structural features can provide information to predict the bandgap. 
        
        If we take a closer look at the \ac{Roost} model, which takes a purely compositional approach to materials property prediction, we can see a poor generalization to the test set (see Table \ref{tab:results:bandgap-2d-tpot-sisso-roost-xgboost}). This indicates that we have likely caused it to over-fit (which could have been improved for example through the use of early stopping). Overall, we may also use this result to further show the need for structural descriptors when predicting bandgaps of these systems.
        
        The \ac{TPOT}, \ac{SISSO}, and \ac{XGBoost} models for 2D material bandgap achieved similar performance to one-another (see Table \ref{fig:results:bandgap-2d-tpot-sisso-roost-xgboost}). Moreover, we again have the opportunity to extract meaning from the \ac{TPOT} model here, as it leveraged Elastic Net to perform its prediction --- a model which is achieved by mixing together Ridge Regression and the \ac{LASSO}. Additionally, as the L1/L2 ratio is 0.95, this is nearly entirely L1 regularization (see section \ref{results:2d-material-bandgaps}). As a result of this, and that the data were all scaled such that they ranged between 0 and 1, we can view the the feature coefficients approximately (but not entirely) through the lens of \ac{LASSO}, which does perform feature selection. Doing this, we see that the corresponding bulk 3D-parent material's bandgap (as found on Materials Project) is the most strongly-weighted feature by the elastic net model. This is extremely intuitive as well --- it is reasonable for the bulk bandgap to be correlated with the 2D bandgap.
        
        The \ac{XGBoost} model, however, suggests an entirely different set of descriptors as being ``important.'' Instead, we find that the five most-important features are, in order from greatest to least, the minimum Mendeleev number, average atomic weight, variance of the number of unfilled s-orbitals, average ground-state \ac{DFT} volume, and maximum Cordero\cite{corderoCovalentRadiiRevisited2008} covalent radius. None of these descriptors have a particularly intuitive relationship with the bandgap either. This is a similar result to that found by Rajan et al. \cite{rajanMachineLearningAssistedAccurateBand2018}, who leveraged several models to predict the GW bandgap of MXene systems; in this work, they found that several of the most-important features exhibited only a statistical (i.e. non-intuitive) relationship with the bandgap.
        
        Finally, the \ac{SISSO} model (Equation \ref{eqn:results:bandgap-sisso}) shows an equivalent performance to \ac{TPOT} and \ac{XGBoost}, while using only three primary features. Similar to the \ac{TPOT} model the bandgap of the bulk 3D-parent material is the most important feature, with only minor non-linear contributions from the minimum atomic Van der Waals radius and period. Interestingly, the \ac{SISSO} model not only misses some of the metallic materials, but also incorrectly predicts some materials to metallic in the training set. This suggests the increase simplicity of the \ac{SISSO} models may slightly reduce their reliability.
        
        % Some features that appear in them, however, do have some intuitive connections with the bandgap (e.g. the polarizability is an electronic property). Despite this, we can achieve some level of interpretation from the model based on its features. One feature of particular interest is the product of the variance of the thermal conductivity with the variance of the boiling point, divided by the square of the average boiling point.
        
        % In the case of graphene, another well-known 2D system, the thermal conductivity has been shown to be almost entirely phonon-based\cite{jacimovskiPhononThermalConductivity2015}, and phonon modes typically depend on the strength of bonds in the system. Further, Rajan et al. \cite{rajanMachineLearningAssistedAccurateBand2018} identified that the average boiling point and the melting point standard deviation had relatively high importance in predicting MXene bandgaps. In their work, they attributed this to a correlation between stronger bonding in MXene systems and larger band-gaps. Overall, what this feature appears to be, is one describing the bond strength of the system. 
        Future work on this problem may achieve better performance on the bandgap problem by investigating the bond strengths of the different elements in the system. We also note the very good performance that recent neural network approaches have had on the 3D bandgap problem\cite{choudharyAtomisticLineGraph2021,chenGraphNetworksUniversal2019}, likely due to their representation of the structure of the 3D systems. Similar to how the \ac{Roost} model achieves good success when compositional descriptors are appropriate, we may find good success in leveraging neural network approaches when structural features are required. We note here that Deng et al.\cite{dengXGraphBoostExtractingGraph2021} achieved good results on a variety of molecule properties by incorporating various graph representations from different neural network architectures. Hence, future work in this domain may benefit from the incorporation of the information-dense structural fingerprints that may be obtained from neural network-based approaches.

    \subsection{2D Material Exfoliation Energy}
    \label{discussion:2d-material-exofoliaiton}
        % Cite for 2D material stability \cite{schlederExploringTwoDimensionalMaterials2020}
        
        We observed some of the worst model performance (across all models) in the case of the 2D material exfoliation energy. Despite being a larger dataset than either the perovskite (144 total, 129 in the training set) or bandgap (1,412 total, 1,270 in the training set), the 3,049 entries in the training set (out of 3,388 total) for the exfoliation energy proved insufficient to achieve good results for any of the models. Moreover, the compositional and structural features were not sufficient to adequately describe the system.
        
        Some interpretation can at least be derived from the \ac{XGBoost} and \ac{TPOT} models. The five most-important features according to \ac{XGBoost} (see Supporting Information Figure \ref{sfig:xgboost-feature-importance:exfoliation-xgboost-importances}) are, in order, the maximum Mendeleev number, average melting point, average evaporation heat, maximum number of unfilled orbitals, and the sum of the melting points. In the case of the \ac{TPOT} model, we arrived at an extremely randomized tree approach, which also has a feature importance metric. Here, we find that the average Van der Waals radius, maximum dipole polarizability, minimum atomic weight, maximum atomic weight, and maximum elemental (and tabulated) \ac{DFT} bandgap are weighted as important. Between the two models, we see much difference in which features are deemed important. In the \ac{XGBoost} model, the average melting point, sum of melting points, and average evaporation heat of the elemental systems can be rationalized if we realize that these are all driven by the strength of the interaction between atoms in the material, thus these descriptors may provide information relating to the forces that must be overcome when exfoliation is performed. Many of the other features in the two models correlate with size: the maximum Mendeleev number, average Van der Waals radius, and minimum / maximum atomic weights all provide a description of the size of the atoms in the system. 
        
        Finally, the \ac{SISSO} model (Equation \ref{eqn:results:exfoliation-sisso-rung2-2term}) echos these findings. Although it is less performant, it provides intuitive descriptors that tell a similar story. One term takes the ratio of the decomposition energy and approximations to the atomic volume. This captures a description of atomic size as well as the strength of atomic bonds. Additionally, the second term in the model incorporates information about the surface, thermal conductivity and the heat or evaporation. Although the second term is less descriptive, it still captures terms relating to the size of the atoms and bond strengths of that atoms involved in the 2D material. The relatively poor performance of the \ac{SISSO} models, also highlights the need for better input features to describe the exfoliation energy of a material. In cases where only a loose correlation between a target property and the inputs exist, the relative simplicity of symbolic regression models can be detrimental. While \ac{TPOT} and \ac{XGBoost} can utilize information from all features in their final predictions, symbolic regression in general and \ac{SISSO} in particular only acts on the order of tens of features. Because of this, it is likely that more structural information is needed to get accurate models with \ac{SISSO}.
        
        In effect, the models can be interpreted as collectively describing the exfoliation energy as a function of i) the size of the atoms in the 2D material and ii) the strength of the intermolecular forces between those atoms. When we predict exfoliation energies, we're predicting the interaction between layers in an exfoliable material. Overall, finding better methods of cheaply approximating these weak interactions may provide better results in the prediction of exfoliation. Additionally, as the number of datasets which contain exfoliation energies increases (such as the \ac{2DMatPedia}\cite{zhou2DMatPediaOpenComputational2019}, \ac{C2DB}\cite{gjerdingRecentProgressComputational2021, haastrupComputational2DMaterials2018} and \ac{JARVIS}\cite{choudharyJointAutomatedRepository2020}), further insight into this problem will be possible, and more-complex (albeit less-interpretible) models will become feasible. 
        
\begin{figure}[H]
            \centering
            \includegraphics[width=\textwidth,height=\textheight,keepaspectratio]{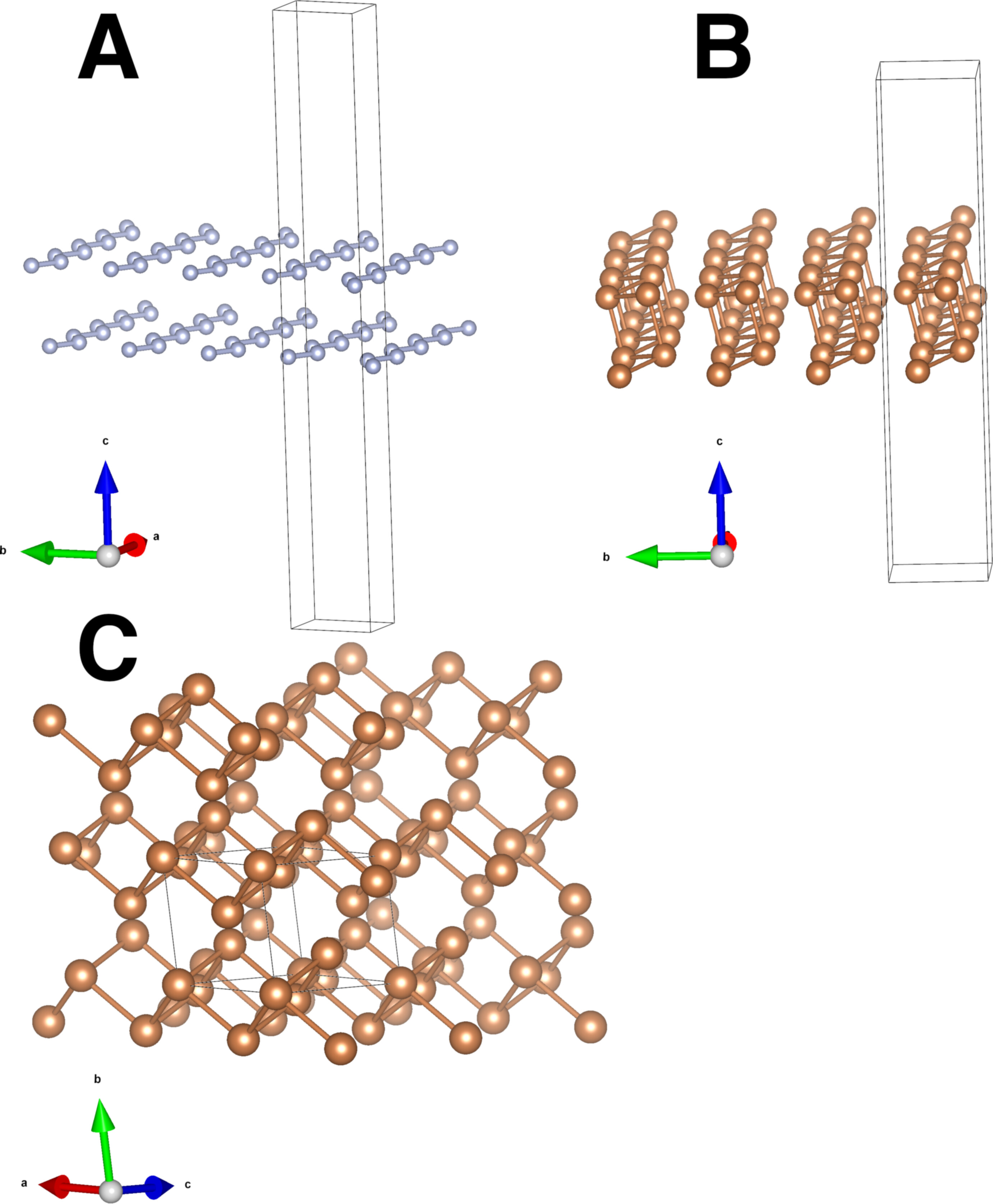}
            \caption{Illustrations of of A) a N-containing system (2dm-id 5985) which persisted as a large outlier across all exfoliation models in the training set, B) the Sb structure (2dm-id 4275) the N-containing system was derived from, and C) the bulk structure from Materials Project (mp-567409) from-which the exfoliation of the Sb system was simulated.}
            \label{fig:discussion:2d-material-exfoliation:outlier}
        \end{figure}

        Additionally, in order to obtain more-accurate predictions of exfoliation energy, data generated via a more thorough computational treatment may be required. We illustrate this by examining an outlier in the training set at 9.9 J / m\textsuperscript{2}, which all four models heavily under-predicted (see Supporting Information figure \ref{sfig:exfoliation-energy-comparison:exfoliation-2d-tpot-sisso-roost-xgboost}) (7-8 eV in the case of \ac{XGBoost} and \ac{TPOT}, and over 9 eV in the case of \ac{Roost} and \ac{SISSO}). Upon closer examination of this system, we find that it is actually a pair of layers containing N atoms (Figure \ref{fig:discussion:2d-material-exfoliation:outlier} A). The \ac{2DMatPedia} \cite{zhou2DMatPediaOpenComputational2019} reports that this system (2dm-id 5985) was not directly sourced by a simulated exfoliation from a bulk structure, but instead was obtained by substituting the atoms in a hypothetical 2D Sb structure (Figure \ref{fig:discussion:2d-material-exfoliation:outlier} B). The Sb structure (2dm-id 4275) was obtained by a simulated exfoliation from a structure obtained from materials project (Figure \ref{fig:discussion:2d-material-exfoliation:outlier} C). The parent bulk material (mp-567409) is reported by the Materials Project\cite{jainCommentaryMaterialsProject2013} to be a monoclinic crystal which undergoes a favorable decomposition (energy above hull is reported as 0.121 eV / atom) to a triclinic system. That being said, as this is a hypothetical 2D system, comparison with the hypothetical 3D bulk system was necessary for the calculation of exfoliation energy. As the prediction of crystal structure is a very challenging field with few easy approximations\cite{oganovModernMethodsCrystal2011}, this may have contributed further to the extreme value of the exfoliation energy. Indeed, as Zhou et al report\cite{zhou2DMatPediaOpenComputational2019}, the decomposition energy lends itself better to assessing whether a material is truly stable. Indeed, despite the extremely high exfoliation energy of this hypothetical 2D N system, it is reported by the \ac{2DMatPedia} to have a decomposition energy of 0 eV/atom. This too seems somehwat high, as systems containing N-N bonds tend to be high-energy materials, typically undergoing strongly exothermic decomposition to inert, gaseous N\textsubscript{2} \cite{kumarExplosiveChemistryNitrogen2019}. With this in conjunction with the observation that our models all predict exfoliation energies significantly lower than the tabulated values, we have reason to believe that this system would be far easier to exfoliate than the ~10 eV exfoliation energy implies. Moreover, this system may have a strong energetic preference to decompose further into N\textsubscript{2}, which additional \ac{DFT} work could reveal. Overall, this underscores the importance of obtaining high-quality data, and filtering that high-quality data, for the training of interpretable models.
        
    \subsection{Future Outlook}
    \label{discussion:future-outlook}
    
        As \ac{ML} is further integrated into materials discovery workflows, we anticipate that the numerous successes neural networks have presented\cite{behlerFourGenerationsHighDimensional2021, yaoTensorMol0ModelChemistry2018, behlerGeneralizedNeuralNetworkRepresentation2007, xieCrystalGraphConvolutional2018, goodallPredictingMaterialsProperties2020, westermayrCombiningSchNetSHARC2020, schuttSchNetContinuousfilterConvolutional2017, toniatoUnassistedNoiseReduction2021, schwallerPredictingRetrosyntheticPathways2020, schwallerMolecularTransformerModel2019, schwallerFoundTranslationPredicting2018, vaucherInferringExperimentalProcedures2021, materDeepLearningChemistry2019, panteleevRecentApplicationsMachine2018, draxlBigDataDrivenMaterials2020, butlerMachineLearningMolecular2018} will continue to propel them onto the cutting edge of chemical property prediction. This comes with the challenge of honing our techniques for their interpretation, an area which has seen much interest in recent years, and where there is still plenty of opportunity for further development\cite{fanInterpretabilityArtificialNeural2021,zhangSurveyNeuralNetwork2021}.
        We also expect \ac{AutoML} techniques such as \ac{TPOT} will continue gaining traction in materials discovery, due to the amount of success and attention they have recently had \cite{gijsbersOpenSourceAutoML2019, yaoTakingHumanOut2019, heAutoMLSurveyStateoftheart2021, leScalingTreebasedAutomated2020, olsonAutomatingBiomedicalData2016, olsonEvaluationTreebasedPipeline2016}. This too presents the challenge of interpretability if highly complex pipelines are generated (see Sections \ref{results:perovskite-volume-prediction} and \ref{discussion:perovskite-volume-per-formula-unit}). We note here that part of the value that \ac{AutoML} techniques bring is the ability to make advanced techniques accessible to a wider audience of researchers by lowering the barrier of entry. Hence, we expect that the problem of interpretation may be compounded for \ac{AutoML} (and especially \ac{NAS}) systems: the ability to automatically extract some level of interpretation from the generated pipelines is important for automation to make \ac{ML} truly accessible to non-experts. Overall, we expect that as neural network models and \ac{AutoML} algorithms continue to grow in capability and complexity, work in developing the tools and techniques needed to interpret them will see a greater attention.
        
        In contrast with challenge of interpreting neural networks or the pipelines found by \ac{AutoML} systems, symbolic regression tools like Eureqa and \ac{SISSO} yield an exact equation describing the model, and are thus easier to interpret. This makes it easier to achieve key insights with physical interpretations --- such as the very intuitive way in which \ac{SISSO} is able to describe the systems. Overall, despite its reduced ability to predict the exfoliation energy of a material when compared to the models of \ac{TPOT}, \ac{XGBoost}, and \ac{Roost}, we note the mathematical equations returned by \ac{SISSO} provide a direct relationship between the target properties and model predictions. Additionally, in the case of the exfoliation energy, we believe that we may see further improvements by including richer structural information. We base this on the observation that the \ac{Roost} model performed poorly on both of this problems -- recalling that \ac{Roost} is only provided the chemical formula of the system, this could indicate that compositional descriptors alone are insufficient to describe these properties. Indeed, it is well-known that structure and energy are intimately related (the fundamental assumption of geometry optimization techniques is that energy is a function of atomic position), hence it can be inferred that exfoliation energy and structure are similarly related. In the case of bandgaps, we note that there is also a strong dependence on structure; Chaves et al\cite{chavesBandgapEngineeringTwodimensional2020} notes that the number of layers in a 2D material can strongly influence the band gap, reporting differences of up to several eV can occur between the bulk and monolayer form of a material.
        
        % \TP{We can consider adding a small section here on how introducing structural primary features may help the overall performance of the SISSO models. This can also help with the discussion where the increased complexity of the other models can help reduce that error.}
        % Added this to the tail end of the preceeding paragraph. Let me know if this doesn't fit well and I can move it. -JD

        Interoperability is still a challenge in the materials discovery ecosystem. Although it is possible to easily convert between different chemical file formats (e.g. by OpenBabel\cite{oboyleOpenBabelOpen2011}), and packages such as Pymatgen \cite{ongPythonMaterialsGenomics2013}, \ac{ASE} \cite{larsenAtomicSimulationEnvironment2017a}, and RDKit \cite{landrumRDKit2021} can easily convert to each others' format, we note that there is a challenge of calculating features using a variety of different packages. Some tools expect Pymatgen objects (e.g. XenonPy), others expect \ac{ASE} objects, whereas others require RDKit objects (e.g. all of the descriptors in the RDKit library) to perform a calculation of features, thus creating some standard for the interoperability of these packages would be beneficial. Additionally, further efforts should be made to report the sources of data used by featurization packages. We note that MatMiner \cite{wardMatminerOpenSource2018} is exemplary in this regard: each of the featurization classes it defines has a ``citation'' method returning the appropriate source to credit. Mendeleev \cite{mentelMendeleevPythonResource2014} is another good example of this; within its documentation, a table lists citations for many (though not all) of the elemental properties it can return. Overall, by placing a stronger focus on i) interoperability and ii) data provenance, the Python materials modeling ecosystem can be made stronger --- and therefore help accelerate materials discovery.
        
        All of the models we have investigated in this work required sufficient training data to avoid over-fitting. Although techniques such as cross-validation, early-stopping (in the case of neural networks and \ac{XGBoost}), and train/test splitting can help guard against (and detect) over-fitting, having a sufficiently-large dataset is of the utmost importance to achieve truly generalizable models. As a result, there is a critical need for data management approaches that satsify the set of \ac{FAIR} principles. This crucial need for effective data management has led to the incorporation of data storage tooling in popular chemistry packages including Pymatgen\cite{ongPythonMaterialsGenomics2013}, \ac{ASE}\cite{larsenAtomicSimulationEnvironment2017a}, and RDKit\cite{landrumRDKit2021}. Moreover, advances in both computational capacity and techniques has given rise to studies performing the high-throughput screening of chemical systems\cite{zhangFirstPrinciplesHighThroughputScreening2020, mayrNovelTrendsHighthroughput2009, deanRapidPredictionBimetallic2020}. This has resulted in the development of tools focusing on the provenance of data, such as the \ac{AiiDA} system \cite{uhrinWorkflowsAiiDAEngineering2021, huberAiiDAScalableComputational2020}.
        
        Overall, we have identified a series of key issues should see more attention as the digital ecosystem surrounding materials modeling continues to develop. First, interpretability of models allows us to derive physical understanding from the available data. This is a key benefit of symbolic regression tools like \ac{SISSO}, which result in the creation of human-readable equations describing the model. Additionally, increasing the accessibility of \ac{ML} techniques through automation (such as in the field of \ac{AutoML}) will allow a wider range of researchers the ability to benefit from advances in modeling techniques. Data management and data provenance is another major issue, which allows us to better understand which datasets can be combined (e.g. when combining \ac{DFT} datasets, the methodologies should be consistent between them), and to help us understand if something intrinsic to the training data is affecting model performance. These data management goals are core focus of platforms such as Exabyte \cite{bazhirovDatacentricOnlineEcosystem2019}, which provides an all-in-one solution for i) storing material data and metadata, ii) storing the methodology required to derive a property from a material, and iii) providing the means to automatically perform calculations, and iv) automatically extracting calculation results and storing them for the user. This focus on providing a tool that manages materials, workflows, and calculations has allowed Exabyte to be a highly successful platform, which has led to studies involving automated phonon calculations\cite{bazhirovFastAccessibleFirstprinciples2018}, high-throughput screening of materials for their band-structure \cite{dasAccessibleComputationalMaterials2018, dasElectronicPropertiesBinary2019}. Future capabilities of the platform are slated to include a categorization scheme for computational models to provide even more metadata to track the provenance of calculated material properties\cite{zechCateComPracticalDatacentric2021}.

\section{Conclusion}
\label{sec:conclusion}

In this work, we have performed a series of benchmarks on a diverse set of \ac{ML} algorithms: gradient boosting (\ac{XGBoost}), \ac{AutoML} (\ac{TPOT}), deep learning (\ac{Roost}), and symbolic regression (\ac{SISSO}). These models were used to predict i) the volume of perovskites, ii) the \ac{DFT} bandgap of 2D materials, and iii) the exfoliation energy of 2D materials. We identify that \ac{TPOT}, \ac{SISSO} and \ac{XGBoost} tend to produce more-accurate models than \ac{Roost},but \ac{Roost} works well in systems where compositional descriptors are enough to predict the target property.
Finally, although \ac{SISSO} was unable to find an accurate model for the exfoliation energy, it provides a human-readable equation describing the model, facilitating an easier interpretation compared to the other algorithms. 
We believe that interpretability will remain a key challenge to address as complex techniques (i.e. neural networks and \ac{AutoML}) become more mainstream within the digital materials modeling ecosystem. Overall, as tools improving the accessibility of machine-learning continue to be developed, data provenance and model interpretability will become even more important, as it is a critical part of ensuring the accessibility of these techniques. By working to ensure that a wider audience of researchers can achieve insight from the rich digital ecosystem of materials design, materials discovery can be accelerated.

\section*{Acknowledgements}
This research was supported by the US \ac{DoE} \ac{SBIR} program (grant no. DE-SC0021514). Computational resources were provided by Exabyte Inc. The authors also wish to acknowledge fruitful discussions with Rhys Goodall (University of Cambridge) regarding the \ac{Roost} framework. 

\section*{Data Access}
Jupyter (Python) notebooks are available on Exabyte's GitHub (\url{https://github.com/Exabyte-io/Scientific-Projects/tree/arXiv_interpretableML_nov2021}), which contains code to reproduce our results and figures.

\section*{Disclosures}
James Dean and Timur Bazhirov are employed by Exabyte Inc.

% --- B I B L I O G R A P H Y ---
\bibliography{refs}

\pagebreak
\appendix
\renewcommand{\thesection}{S}
% Prepend tables and figures with S
\renewcommand{\thetable}{S\arabic{table}}
\renewcommand{\thefigure}{S\arabic{figure}}
\renewcommand{\theequation}{S\arabic{equation}}

% Restart table and figure numbering
\setcounter{table}{0}
\setcounter{figure}{0}
\setcounter{equation}{0}

\section{Supporting Information}
\label{sec:supporting-information}

    \subsection{XGBoost Feature Importance}
    \label{si:xgboost-feature-importance}
        The \ac{XGBoost} approach allows us to assess the feature importance scores of the various model inputs. In this section, we report the importance scores for the 10 most-important features identified by each model. Feature importances are calculated using the default "Gain" metric available in \ac{XGBoost}.
       
        \begin{figure}[H]
            \centering
            \includegraphics[width=\textwidth,height=\textheight,keepaspectratio]{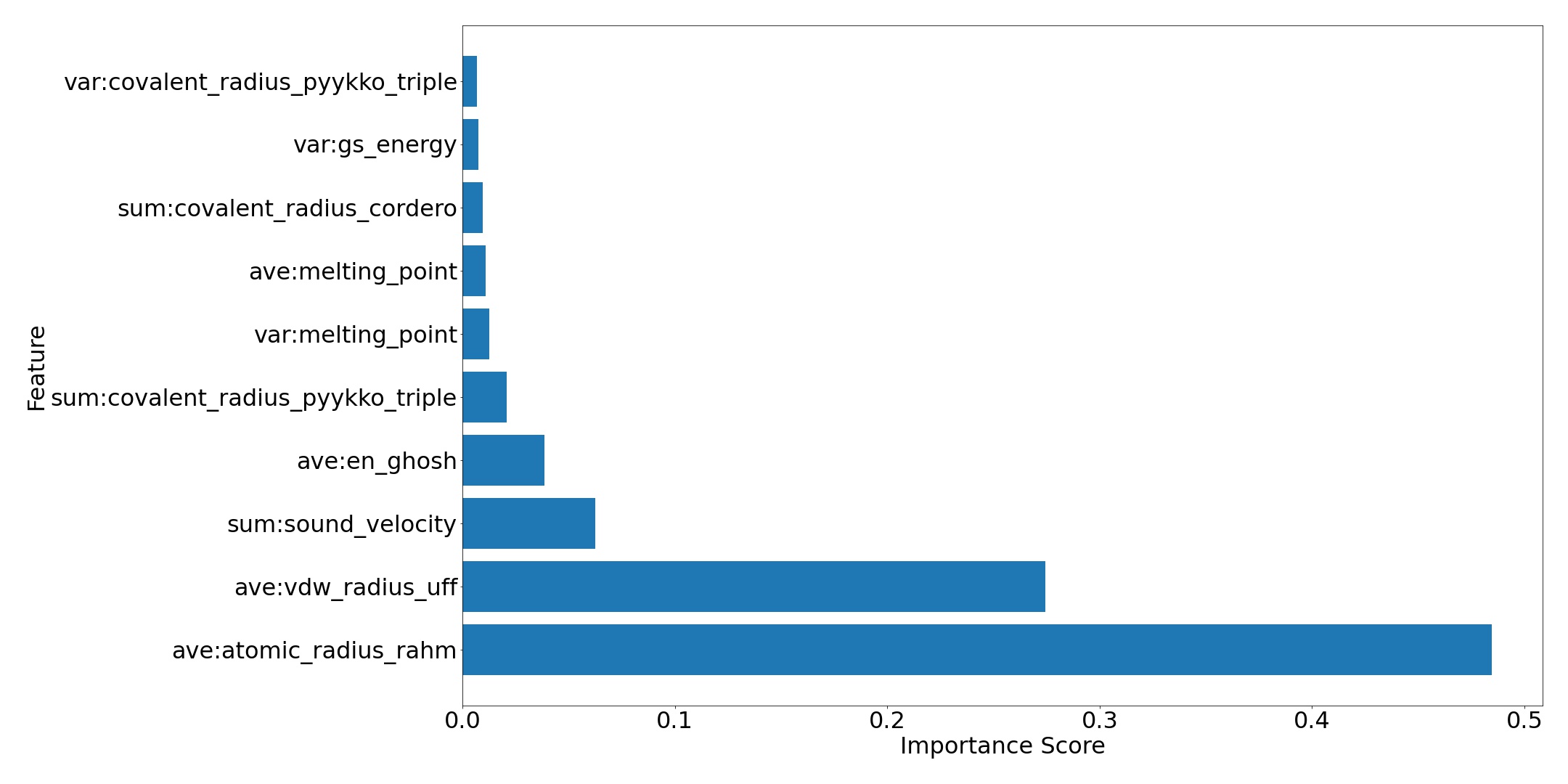}
            \caption{Importance metrics for the 10 most-important features identified by the \ac{XGBoost} perovskite volume model using the ``gain'' feature importance metric.}
            \label{sfig:xgboost-feature-importance:perovskite-xgboost-importances}
        \end{figure}
        
        \begin{figure}[H]
            \centering
            \includegraphics[width=\textwidth,height=\textheight,keepaspectratio]{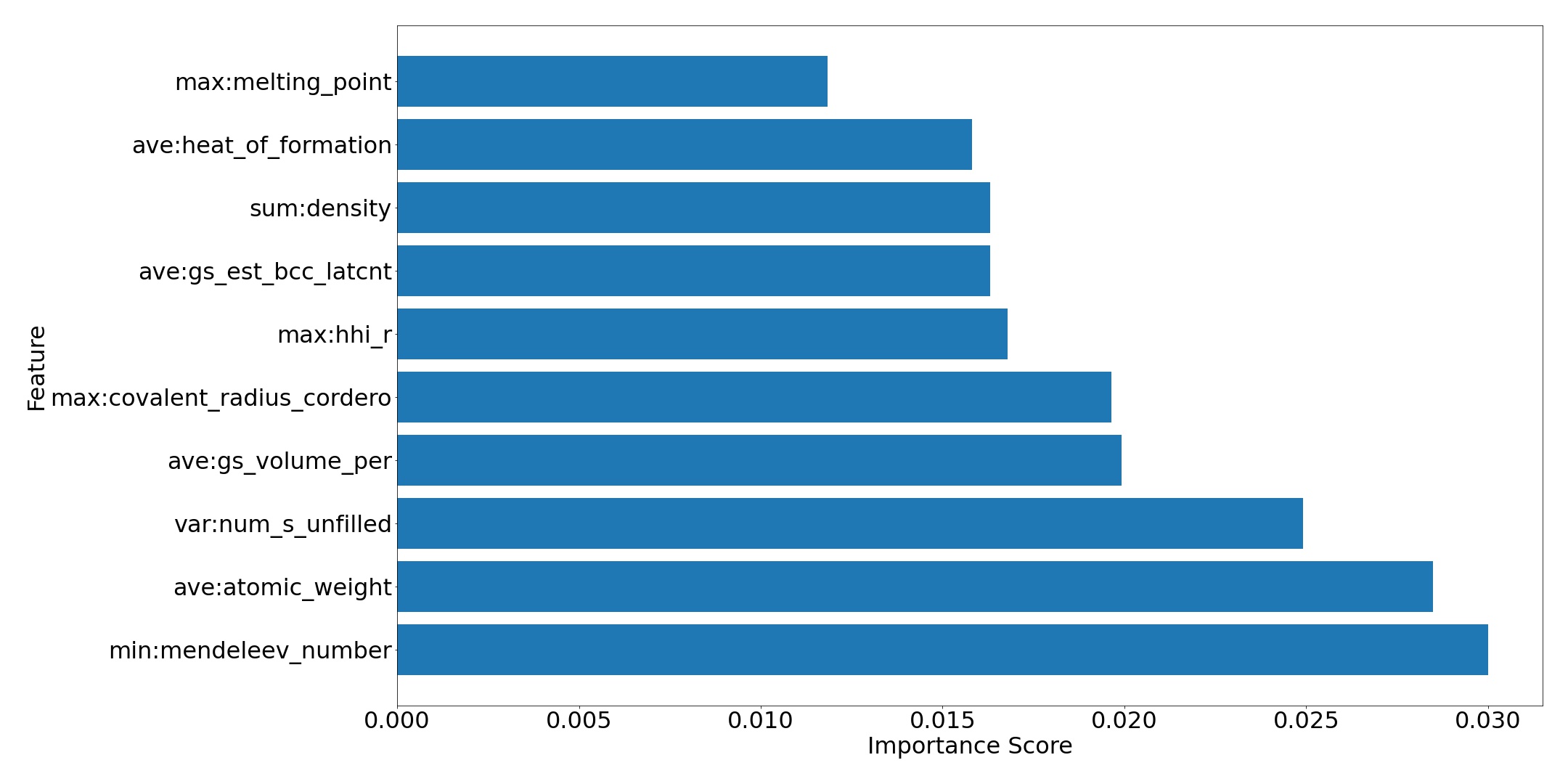}
            \caption{Importance metrics for the 10 most-important features identified by the \ac{XGBoost} 2D material bandgap model using the ``gain'' feature importance metric.}
            \label{sfig:xgboost-feature-importance:bandgap-xgboost-importances}
        \end{figure}
        
        \begin{figure}[H]
            \centering
            \includegraphics[width=\textwidth,height=\textheight,keepaspectratio]{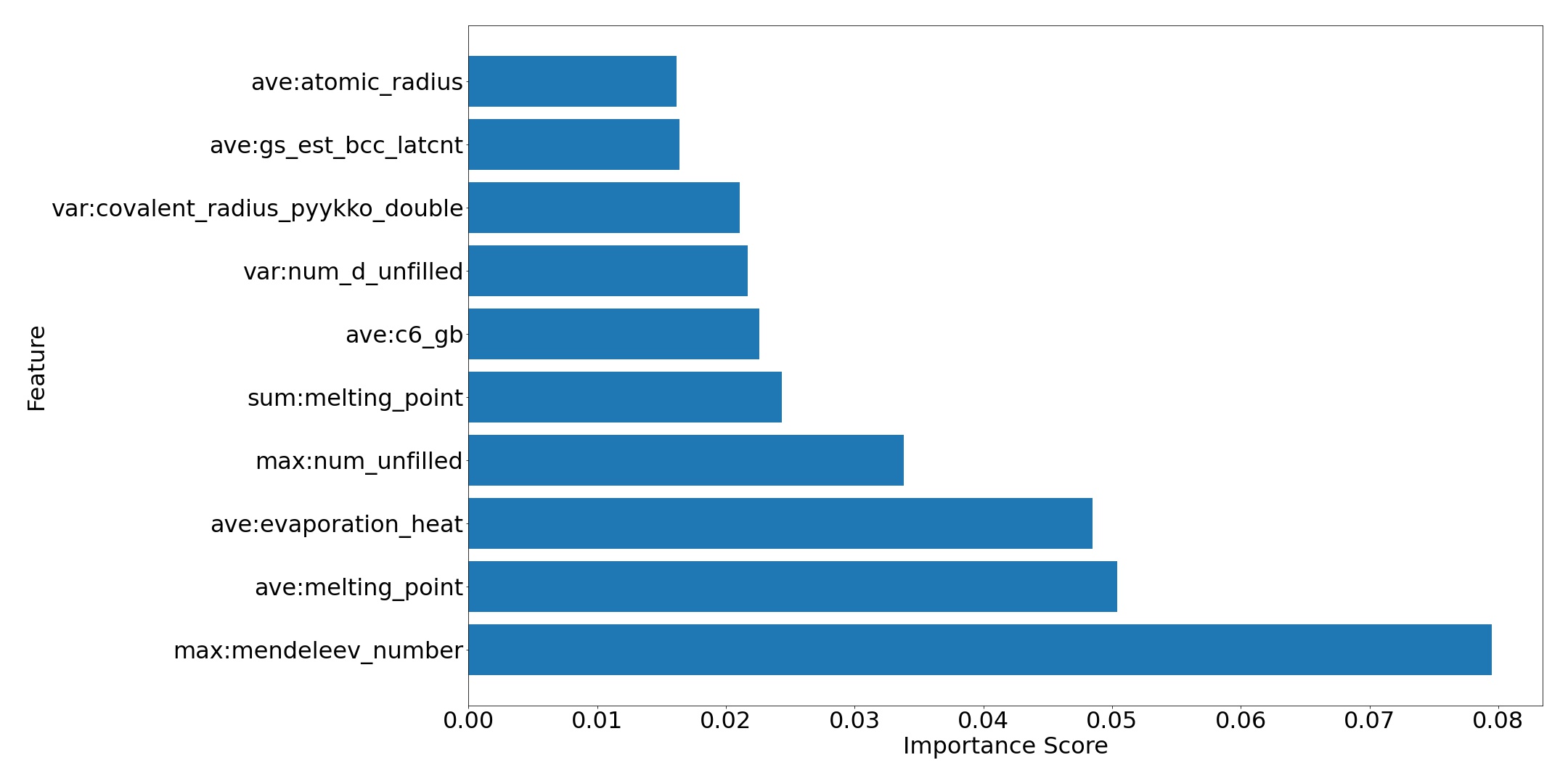}
            \caption{Importance metrics for the 10 most-important features identified by the \ac{XGBoost} 2D material exfoliation model using the ``gain'' feature importance metric.}
            \label{sfig:xgboost-feature-importance:exfoliation-xgboost-importances}
        \end{figure}
        
    \subsection{TPOT Model Components}
    
        To facilitate the reproducibility of our work, we have provided a list of the model components that \ac{TPOT} could combine to generate models in \ref{stab:tpot-model-components:model-selection}. For more information on our specific methodology and other \ac{TPOT} sections, see Section \ref{methodology:ml-models:automl-with-tpot} in the manuscript.
        
        \begin{table}[H]
            \centering
            \rowcolors{2}{gray!25}{white}
            \caption{Search space of model components that are allowed by \ac{TPOT} in its default regression configuration. Components are listed alongside the package they are sourced from, using the same class name defined in their respective package. Additionally, we list the role that the component.}
            \begin{tabular}{rcc}
            \rowcolor{gray!50}
            Model Name & Source Package & Role \\
            \hline
            \texttt{ElasticNetCV} & Scikit-Learn & Estimator \\
            \texttt{ExtraTreesRegressor} & Scikit-Learn & Estimator \\
            \texttt{GradientBoostingRegressor} & Scikit-Learn & Estimator \\
            \texttt{AdaBoostRegressor} & Scikit-Learn & Estimator \\
            \texttt{DecisionTreeRegressor} & Scikit-Learn & Estimator \\
            \texttt{KNeighborsRegressor} & Scikit-Learn & Estimator \\
            \texttt{LassoLarsCV} & Scikit-Learn & Estimator \\
            \texttt{LinearSVR} & Scikit-Learn & Estimator \\
            \texttt{XGBRegressor} & \acs{XGBoost} & Estimator \\
            \texttt{SGDRegressor} & Scikit-Learn & Estimator \\
            \texttt{Binarizer} & Scikit-Learn & Pre-Processing \\
            \texttt{FastICA} & Scikit-Learn & Pre-Processing \\
            \texttt{FeatureAgglomeration} & Scikit-Learn & Pre-Processing \\
            \texttt{MaxAbsScaler} & Scikit-Learn & Pre-Processing \\
            \texttt{MinMaxScaler} & Scikit-Learn & Pre-Processing \\
            \texttt{Normalizer} & Scikit-Learn & Pre-Processing \\
            \texttt{Nystroem} & Scikit-Learn & Pre-Processing \\
            \texttt{PCA} & Scikit-Learn & Pre-Processing \\
            \texttt{PolynomialFeatures} & Scikit-Learn & Pre-Processing \\
            \texttt{RBFSampler} & Scikit-Learn & Pre-Processing \\
            \texttt{RobustScaler} & Scikit-Learn & Pre-Processing \\
            \texttt{StandardScaler} & Scikit-Learn & Pre-Processing \\
            \texttt{ZeroCount} & TPOT & Pre-Processing \\
            \texttt{OneHotEncoder} & TPOT & Pre-Processing \\
            \texttt{SelectFwe} & Scikit-Learn & Feature Selection \\
            \texttt{SelectPercentile} & Scikit-Learn & Feature Selection \\
            \texttt{VarianceThreshold} & Scikit-Learn & Feature Selection \\
            \texttt{SelectFromModel} & Scikit-Learn & Feature Selection \\
            \texttt{StackingEstimator} & TPOT & Meta-Transformer \\
            \end{tabular}
            \label{stab:tpot-model-components:model-selection}
        \end{table}
    
    \subsection{TPOT Feature Importance}
    \label{si:tpot-feature-importance}
\begin{figure}[H]
            \centering
            \includegraphics[width=\textwidth,height=\textheight,keepaspectratio]{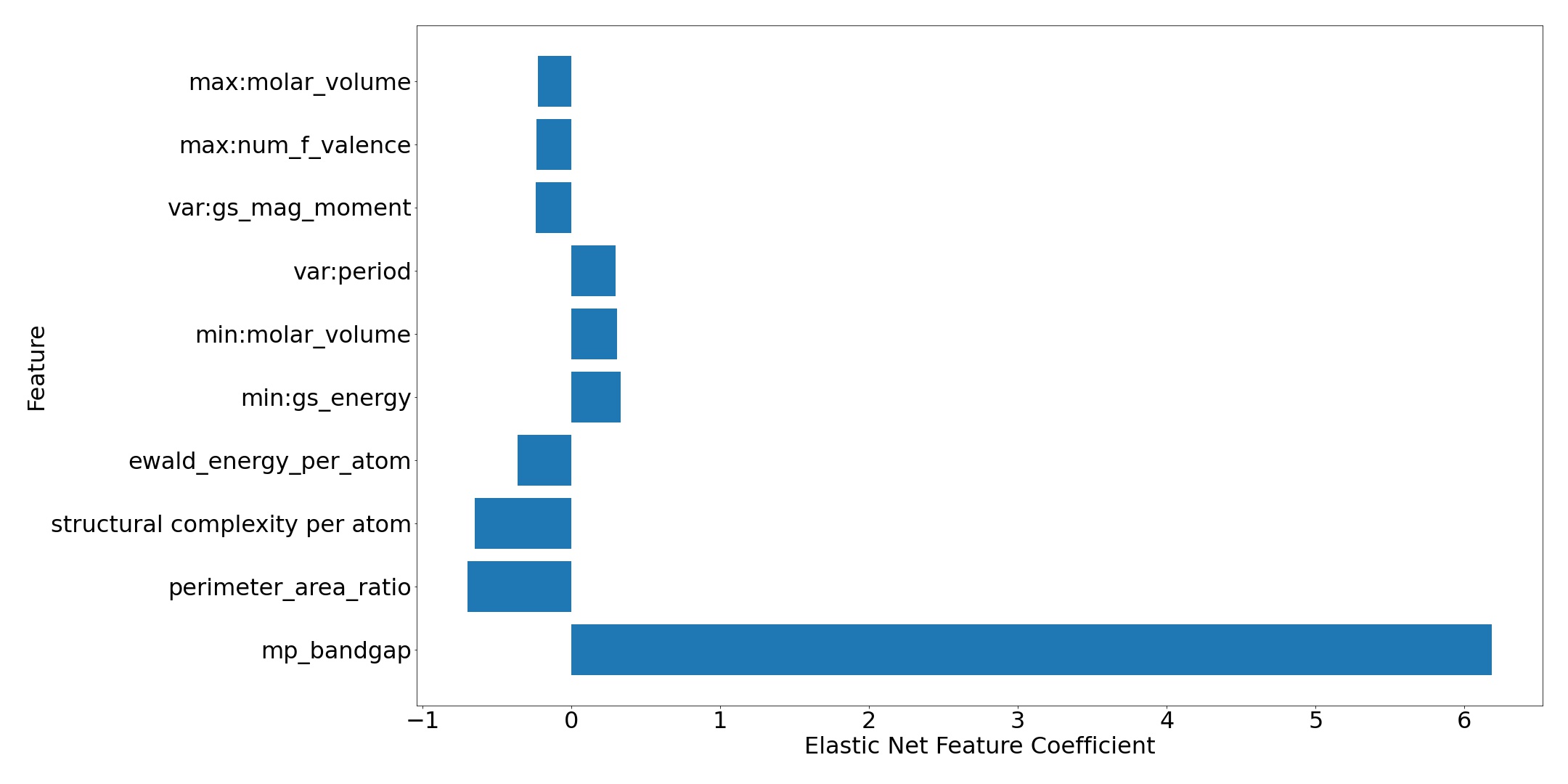}
            \caption{Importance metrics for the 10 most-important features for the 2D material bandgap problem, as identified by the Elastic Net model within the \ac{TPOT} generated pipeline.}
            \label{sfig:tpot-feature-importance:bandgap-tpot-importances}
        \end{figure}

\begin{figure}[H]
            \centering
            \includegraphics[width=\textwidth,height=\textheight,keepaspectratio]{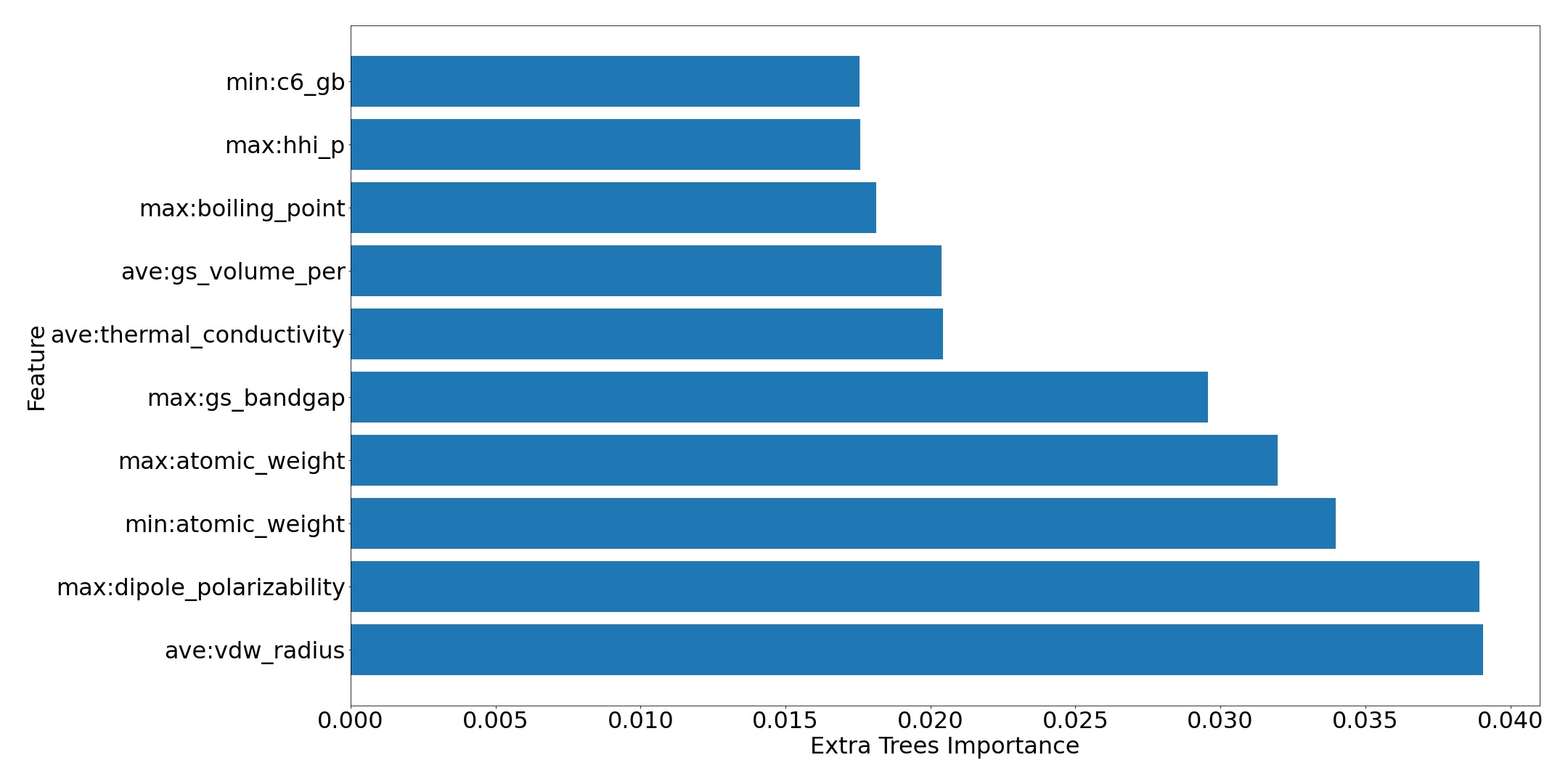}
            \caption{Importance metrics for the 10 most-important features on the 2D material exfoliation energy prediction problem, identified by the Extremely Randomized Trees model within the \ac{TPOT} generated pipeline.}
            \label{sfig:tpot-feature-importance:exfoliation-tpot-importances}
        \end{figure}

        Several of the models generated by \ac{TPOT} can be used to assess how important a variable is. Although the linear support vector machine pipeline does not lend itself well to interpretation in the perovskite volume problem, the models created for the 2D material bandgap (Figure \ref{sfig:tpot-feature-importance:bandgap-tpot-importances}) and the 2D material exfoliation energy (Figure \ref{sfig:tpot-feature-importance:exfoliation-tpot-importances}) lend themselves to some degree of interpretation.
        
    % \subsection{LASSO Feature Selection}
    \subsection{Prescreened features for SISSO}
    \label{si:sisso-feature-selection}
        % \ac{LASSO} was leveraged to conduct feature selection for the \ac{SISSO} models. The results of the feature selection can be found in the tables in this section, for all three of the problems explored in this work. Additionally, we describe the assumptions we made for the units of these descriptors.
        Here we present the set of thirty features used to find the SISSO models for each problem.
        Additionally, we describe the assumptions we made for the units of these descriptors.
        
        \begin{table}[H]
            \centering
            \rowcolors{2}{gray!25}{white}
            \caption{Results feature pre-screening for \ac{SISSO} for the XenonPy compositional descriptors on the perovskite volume prediction problem. Also shown is the assumption we made for units of the XenonPy descriptors (see Section \ref{methodology:ml-models:symbolic-regression-with-sisso} for more details).}
            \begin{tabular}{rcl}
                \rowcolor{gray!50}
                Feature & Units Assumption & Description \\
                \hline
                \texttt{min:c6\_gb}  & $\mathrm{E_h / a_0^6}$ & The min atomic C\_6 dispersion coefficient\\
                \texttt{min:melting\_point}  & K & The min elemental melting point\\
                \texttt{min:Polarizability}  & \AA & The min elemental polarizability\\
                \texttt{min:atomic\_number}  & --- & The min atomic number \\
                \texttt{min:fusion\_enthalpy}  & KJ / mol & The min fusion enthalpy\\
                \texttt{min:dipole\_polarizability}  & \AA$^3$ & The min elemental dipole polarizablity  \\
                \texttt{min:boiling\_point}  & K & The min elemental boiling point\\
                \texttt{min:gs\_volume\_per}  & $\mathrm{cm^3 / mol}$ & The min DFT volume per atom  \\
                \texttt{min:covalent\_radius\_pyykko\_triple}  & pm & The min single bond covalent radius\\
                \texttt{var:covalent\_radius\_pyykko\_triple}  & pm & The var of the triple bond covalent radius \\
                \texttt{min:covalent\_radius\_slater}  & pm & The min covalent radius by Slater\\
                \texttt{min:covalent\_radius\_pyykko\_double}  & pm & The min double bond covalent radius\\
                \texttt{min:period}  & --- & The min period\\
                \texttt{min:covalent\_radius\_pyykko}  & pm & The min double bond covalent radius\\
                \texttt{min:covalent\_radius\_cordero}  & pm & The min covalent radius by Cordero et al \\
                \texttt{min:atomic\_weight}  & Da & The min atomic weight\\
                \texttt{min:hhi\_p}  & --- & The min \ac{HHI} production values\\
                \texttt{max:en\_allen}  & eV & The max electronegativity by Allen et al\\
                \texttt{ave:covalent\_radius\_pyykko\_triple}  & pm & The mean triple bond covalent radius\\
                \texttt{min:atomic\_radius\_rahm}  & pm & The min atomic radius by Rahm et al\\
                \texttt{ave:atomic\_number}  & --- & The mean atomic number \\ 
                \texttt{ave:gs\_volume\_per}  & $\mathrm{cm^3 / mol}$ & The mean DFT volume per atom \\
                \texttt{ave:heat\_capacity\_mass}  & J / g / K & The mean mass specific heat capacity \\
                \texttt{ave:covalent\_radius\_slater}  & pm & The mean covalent radius by Slater\\
                \texttt{ave:covalent\_radius\_pyykko}  & pm & The mean single bond covalent radius\\
                \texttt{max:mendeleev\_number}  & --- & The max Mendeleev number\\
                \texttt{min:vdw\_radius\_alvarez}  & pm & The min vdw radius by Alvarez\\
                \texttt{min:vdw\_radius\_mm3}  & pm & The min vdw radius according to the MM3 FF\\
                \texttt{max:en\_ghosh}  & EN & The max electronegativity by Ghosh et al.\\
                \texttt{min:vdw\_radius}  & pm & The min vdw radius\\
            \end{tabular}
            \label{stab:perovskite-sisso-importances}
        \end{table}
    
        \begin{table}[H]
            \centering
            \rowcolors{2}{gray!25}{white}
            \caption{Results feature pre-screening for \ac{SISSO} for the compositional and structural descriptors defined in Section \ref{methodology:feature-engineering}, for the 2D material bandgap prediction problem. Also shown is the assumption we made for units of the XenonPy descriptors (see Section \ref{methodology:ml-models:symbolic-regression-with-sisso} for more details).}
            \begin{tabular}{rcl}
                \rowcolor{gray!50}
                Feature & Units Assumption & Description \\
                \hline
                \texttt{mp\_bandgap} & eV & \\
                \texttt{min:c6\_gb} & $\mathrm{E_h / a_0^6}$ & The min atomic C\_6 dispersion coefficient \\
                \texttt{min:covalent\_radius\_slater} & pm & The min covalent radius by Slater\\
                \texttt{min:boiling\_point} & K & The min boiling point \\
                \texttt{min:covalent\_radius\_cordero} & pm & The min covalent radius by Cordero et al\\
                \texttt{min:melting\_point} & K & The min melting point \\
                \texttt{min:dipole\_polarizability} & \AA$^3$ & The min elemental dipole polarizablity \\
                \texttt{min:covalent\_radius\_pyykko} & pm & The min single bond covalent radius\\
                \texttt{min:Polarizability} & \AA$^3$ & The min elemental polarizablity\\
                \texttt{ave:boiling\_point} & K & The mean boiling point \\
                \texttt{min:period} & --- & The min period \\
                \texttt{ave:density} & $\mathrm{g / cm^3}$ & The mean density \\
                \texttt{ave:covalent\_radius\_cordero} & pm & The mean covalent radius by Cordero et al\\
                \texttt{ave:covalent\_radius\_slater} & pm & The mean covalent radius by Slater\\
                \texttt{max:en\_allen} & eV & The max electronegativity according to Allen \\
                \texttt{sum:first\_ion\_en} & eV & The sum of the first ionization energies\\
                \texttt{min:vdw\_radius\_alvarez} & pm & The min vdw radius by Alvarez \\
                \texttt{ave:covalent\_radius\_pyykko} & pm & The mean single bond covalent radius\\
                \texttt{ave:period} & --- & The mean period \\
                \texttt{ave:melting\_point} & K & The mean melting point \\
                \texttt{min:vdw\_radius} & pm & the min vdw radius \\
                \texttt{max:mendeleev\_number} & --- & The max Mendeleev number\\
                \texttt{min:covalent\_radius\_pyykko\_triple} & pm & The min triple bond covalent radius\\
                \texttt{min:covalent\_radius\_pyykko\_double} & pm & The min double bond covalent radius\\
                \texttt{max:first\_ion\_en} & eV & The max first ionization energy\\
                \texttt{min:atomic\_radius\_rahm} & pm & The min atomic radius by Rahm et al\\
                \texttt{sum:specific\_heat} & J / g / K & The sum of the specfic heats at 293 K \\
                \texttt{min:vdw\_radius\_mm3} & pm & The min vdw radius according to the MM3 FF\\
                \texttt{sum:en\_allen} & eV & The sum of the Allen electronegativities \\
                \texttt{max:en\_pauling} & EN & The max of the Pauling electronegativities \\
            \end{tabular}
            \label{stab:bandgap-sisso-importances}
        \end{table}
        
        \begin{table}[H]
            \centering
            \rowcolors{2}{gray!25}{white}
            \caption{Results feature pre-screening for \ac{SISSO} for the compositional and structural descriptors defined in Section \ref{methodology:feature-engineering}, for the 2D material exfoliation energy prediction problem. Also shown is the assumption we made for units of the XenonPy descriptors (see Section \ref{methodology:ml-models:symbolic-regression-with-sisso} for more details).}
            \begin{tabular}{rcl}
                \rowcolor{gray!50}
                Feature & Units Assumption & Description \\
                \hline
                \texttt{min:bulk\_modulus} & GPa & The min bulk modulus \\
                \texttt{min:thermal\_conductivity} & W / m / K & The min thermal conductivity \\
                \texttt{decomposition\_energy} & eV/atom & The energy of decomposition of the material\\
                \texttt{min:evaporation\_heat} & KJ / mol & The min evaporation heat \\
                \texttt{min:boiling\_point} & K & The min boiling point\\
                \texttt{ave:electron\_affinity} & eV & The mean electron affinity\\
                \texttt{ave:thermal\_conductivity} & W / m / K & The mean thermal conductivity\\
                \texttt{max:electron\_affinity} & eV & The max electron affinity\\
                \texttt{var:heat\_capacity\_molar} & J / mol / K & The var in the heat capacity \\
                \texttt{min:melting\_point} & K & The min melting point \\
                \texttt{min:heat\_of\_formation} & KJ / mol & The mininum heat of Formation \\
                \texttt{max:heat\_capacity\_molar} & J / mol / K & The max heat molar capacity\\
                \texttt{ave:icsd\_volume} & $\mathrm{cm^3}$ / mol & The mean ICSD volume \\
                \texttt{sum:thermal\_conductivity} & W / m / K & The sum of all of the thermal conductivities \\
                \texttt{bandgap} & eV & The band gap \\
                \texttt{min:vdw\_radius\_uff} & pm & The min vdw radius from the UFF \\
                \texttt{max:thermal\_conductivity} & W / m / K & The max thermal conductivity \\
                \texttt{ave:heat\_capacity\_molar} & J / mol / K & The mean molar heat capacity\\
                \texttt{var:electron\_affinity} & eV & The varaince in electron affinity\\
                \texttt{ave:vdw\_radius\_uff} & pm &  The mean vdw radius from the UFF \\
                \texttt{perimeter\_area\_ratio} & \AA$^{-1}$ & The ratio between the perimeter and area of the surface \\
                \texttt{ave:bulk\_modulus} & GPa & The mean bulk modulus \\
                \texttt{ave:atomic\_volume} & $\mathrm{cm^3 / mol}$ & The mean atomic volume \\
                \texttt{max:icsd\_volume} & $\mathrm{cm^3 / mol}$ & The max ICSD atomic volume\\
                \texttt{ave:evaporation\_heat} & KJ / mol & The avearge evaporation heat \\
                \texttt{ave:mendeleev\_number} & --- & The mean Mendeleev number \\
                \texttt{ave:hhi\_r} & --- & The mean \ac{HHI} reserve values \\
                \texttt{ave:num\_p\_valence} & --- & The mean number of valance p electrons \\
                \texttt{max:num\_p\_valence} & --- & The max number of valance p electrons \\
                \texttt{var:num\_p\_valence} & --- & The var of number of valance p electrons \\
            \end{tabular}
            \label{stab:exfoliation-sisso-importances}
        \end{table}

    \subsection{Exfoliation Energy Comparison}
    \label{si:exfoliation-energy-comparison}
\begin{figure}[H]
            \centering
            \includegraphics[width=\textwidth,height=\textheight,keepaspectratio]{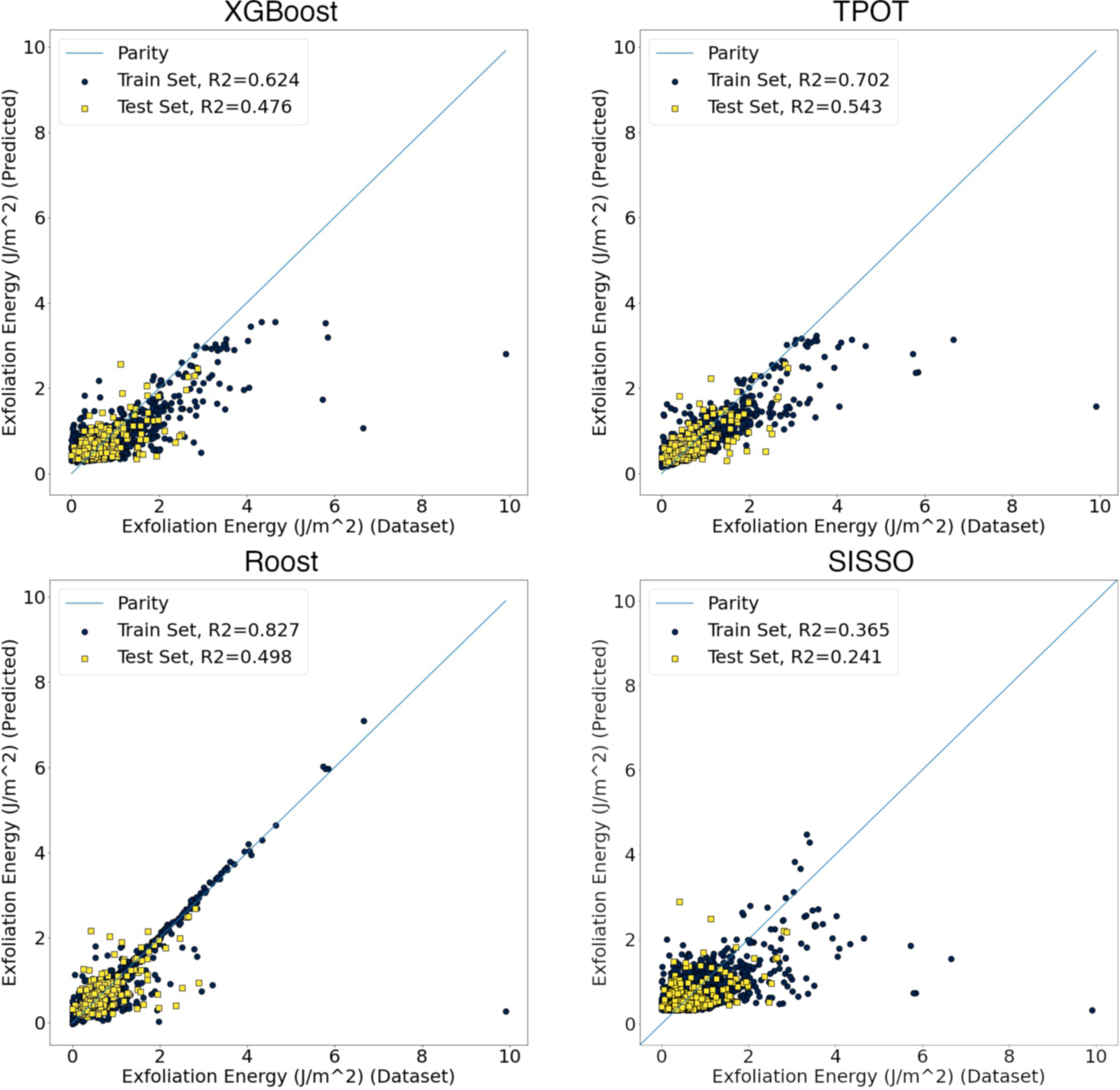}
            \caption{Parity plots for the \ac{XGBoost}, \ac{TPOT}, \ac{Roost}, and \ac{SISSO} models on the 2D  material exfoliation energy problem. Included are the training and testing sets. A diagonal line indicating parity is drawn as a guide to the eye. The SISSO model we report is the rung 2, 2-term model. Regression statistics for the models shown on this plot can be found in Table \ref{tab:results:exfoliation-2d-tpot-sisso-roost-xgboost}.}
            \label{sfig:exfoliation-energy-comparison:exfoliation-2d-tpot-sisso-roost-xgboost}
        \end{figure}

        To facilitate comparison at more-relevant exfoliation energies, we zoomed in the plot featured in Figure \ref{fig:results:exfoliation-2d-tpot-sisso-roost-xgboost} to range up to 2 eV. We present the full parity plot in this section, in Figure \ref{sfig:exfoliation-energy-comparison:exfoliation-2d-tpot-sisso-roost-xgboost}.
    
    % ============================
    % XGBoost Metal Classification
    % ============================
    \subsection{Metal Classification with XGBoost}
    \label{si:metal-classification-with-xgboost}
    
        \subsubsection{Structural Fingerprints}
        \label{si:metal-classification-with-xgboost:structural-fingerprints}

             For the purposes of 2D material metallicity classification, we leveraged the the Sine Matrix \cite{faberCrystalStructureRepresentations2015} fingerprint as implemented by the DScribe package \cite{himanenDScribeLibraryDescriptors2020}. The intuition of this fingerprint is that, although it has little to no physical interpretation, it is periodic unlike the Coulomb Matrix \cite{ruppFastAccurateModeling2012}, and captures the infinite energy resulting from two atoms overlapping \cite{himanenDScribeLibraryDescriptors2020}. The descriptor creates an \(N \times N\) matrix, where \(N\) is the number of atoms in the cell. We then take the eigenspectrum (i.e. the eigenvalues sorted from largest to smallest) of this matrix to generate a feature vector of length N. Because the 2D structures considered for the metal classifier have between 1 and 40 atoms within the unit cell, we 0-pad the eigenspectrum such that it is always a vector of length 40.
             
             Like the other models in this study, 10\% of the data was held out as a testing set. While performing the train/test split, we stratify the data to ensure the same proportion of metal / nonmetal data is in the training and testing set. Finally, in order to we leveraged K-Means \ac{SMOTE} as implemented by Imbalanced-Learn \cite{JMLR:v18:16-365}, with K set to 4 neighbors.
    
        \subsubsection{Classification Methodology}\mbox{}
        \label{si:metal-classification-with-xgboost:classification_methodology}
            
            The Sine Matrix Eigenspectrum described in Section \ref{si:metal-classification-with-xgboost:structural-fingerprints} was used as the input feature to predict whether a material was a metal or nonmetal. 
            
            Early stopping was more-aggressively applied, such that if 5 rounds passed without improvement in the \ac{ROC} \ac{AUC} of the validation set, \ac{XGBoost} halted further training. Multiple evaluation metrics were considered in the training process. \ac{XGBoost} was set to internally optimize the logistic regression score of the model, and Optuna was set to optimize the F1 score of the model, as implemented in SciKit-Learn \cite{scikit-learn}.
                
        \subsubsection{Regression Methodology}\mbox{}
        \label{si:metal-classification-with-xgboost:regression_methodology}
        
            An \ac{XGBoost} model was trained using the metal/nonmetal predictions of the Metallicity Classifier described in section \ref{si:metal-classification-with-xgboost:classification_methodology}. The same training and testing set used for the metal classifier was applied here as well, with all systems predicted by the classifier to be metals removed. Features used included the XenonPy descriptors described in section \ref{methodology:feature-engineering:compositional-descriptors} and the structural descriptors from section \ref{methodology:feature-engineering:structural-descriptors:matminer-descriptors}. Because the data being fed into this model was roughly Poisson-distributed, we applied Poisson-based error metrics. \ac{XGBoost} was set to optimize the squared error, and Optuna was set to minimize the mean Poisson deviance. The \ac{TPE} sampler with Hyperband pruning was again applied here. For \ac{XGBoost} predictions on the perovskite volume, we used only the compositional features from XenonPy.
            
    \label{supporting-information:metal-classification-with-xgboost}
        Confusion matrices to summarize the model's performance on the training and test sets can be found in Table \ref{stab:xgboost-metal-classifier-training} and Table \ref{stab:xgboost-metal-classifier-test} respectively.
        
        \begin{table}[H]
            \centering
            \caption{Confusion matrix for the XGBoost Metal Classification training set}
            \begin{tabular}{cccc}
                \multicolumn{1}{c}{} &\multicolumn{1}{c}{} &\multicolumn{2}{c}{Predicted} \\ 
                \multicolumn{1}{c}{} & 
                \multicolumn{1}{c}{} & 
                \multicolumn{1}{c}{Nonmetal} & 
                \multicolumn{1}{c}{Metal} \\ 
                \multirow[c]{2}{*}{\rotatebox[origin=tr]{90}{Actual}}
                & Nonmetal  & 2550  & 333   \\[1.5ex]
                & Metal     & 353   & 2479 \\
            \end{tabular}
            \label{stab:xgboost-metal-classifier-training}
        \end{table}
        
        \begin{table}[H]
            \centering
            \caption{Confusion matrix for the XGBoost Metal Classification test set}
            \begin{tabular}{cccc}
                \multicolumn{1}{c}{} &\multicolumn{1}{c}{} &\multicolumn{2}{c}{Predicted} \\ 
                \multicolumn{1}{c}{} & 
                \multicolumn{1}{c}{} & 
                \multicolumn{1}{c}{Nonmetal} & 
                \multicolumn{1}{c}{Metal} \\ 
                \multirow[c]{2}{*}{\rotatebox[origin=tr]{90}{Actual}}
                & Nonmetal  & 248  & 73   \\[1.5ex]
                & Metal     & 92   & 223  \\
            \end{tabular}
            \label{stab:xgboost-metal-classifier-test}
        \end{table}
        
        Metrics summarizing model performance can be found in Table \ref{stab:xgboost-metal-classifier-metrics}.
        
        \begin{table}[H]
            \centering
            \rowcolors{2}{gray!25}{white}
            \caption{Model performance metrics for the XGBoost Metal Classifier}
            \begin{tabular}{rcc}
                \rowcolor{gray!50}
                Metric            & Training Set & Test Set \\
                \ac{TPR}          & 0.875        & 0.708 \\
                \ac{FPR}          & 0.116        & 0.227 \\
                Accuracy          & 0.880        & 0.741 \\
                F1 Score          & 0.878        & 0.730 \\
                \ac{ROC} \ac{AUC} & 0.947        & 0.821
            \end{tabular}
            \label{stab:xgboost-metal-classifier-metrics}
        \end{table}
        
\begin{figure}[H]
            \centering
            \includegraphics[width=\textwidth,height=\textheight,keepaspectratio]{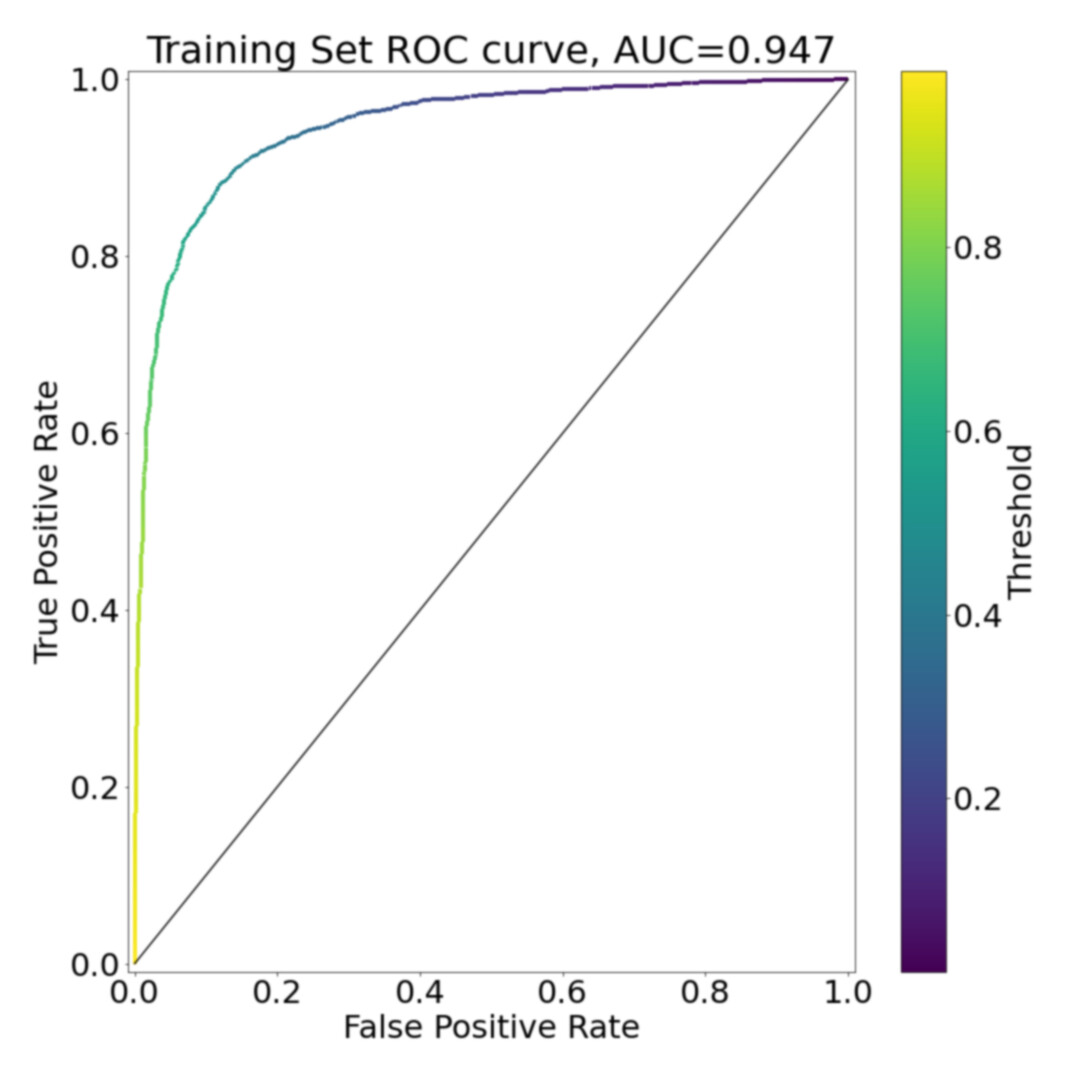}
            \caption{Training-Set \ac{ROC} curve for \ac{XGBoost} model classifying whether a material is metallic or not.}
            \label{sfig:xgboost-metal-classifier-roc-train}
        \end{figure}

\begin{figure}[H]
            \centering
            \includegraphics[width=\textwidth,height=\textheight,keepaspectratio]{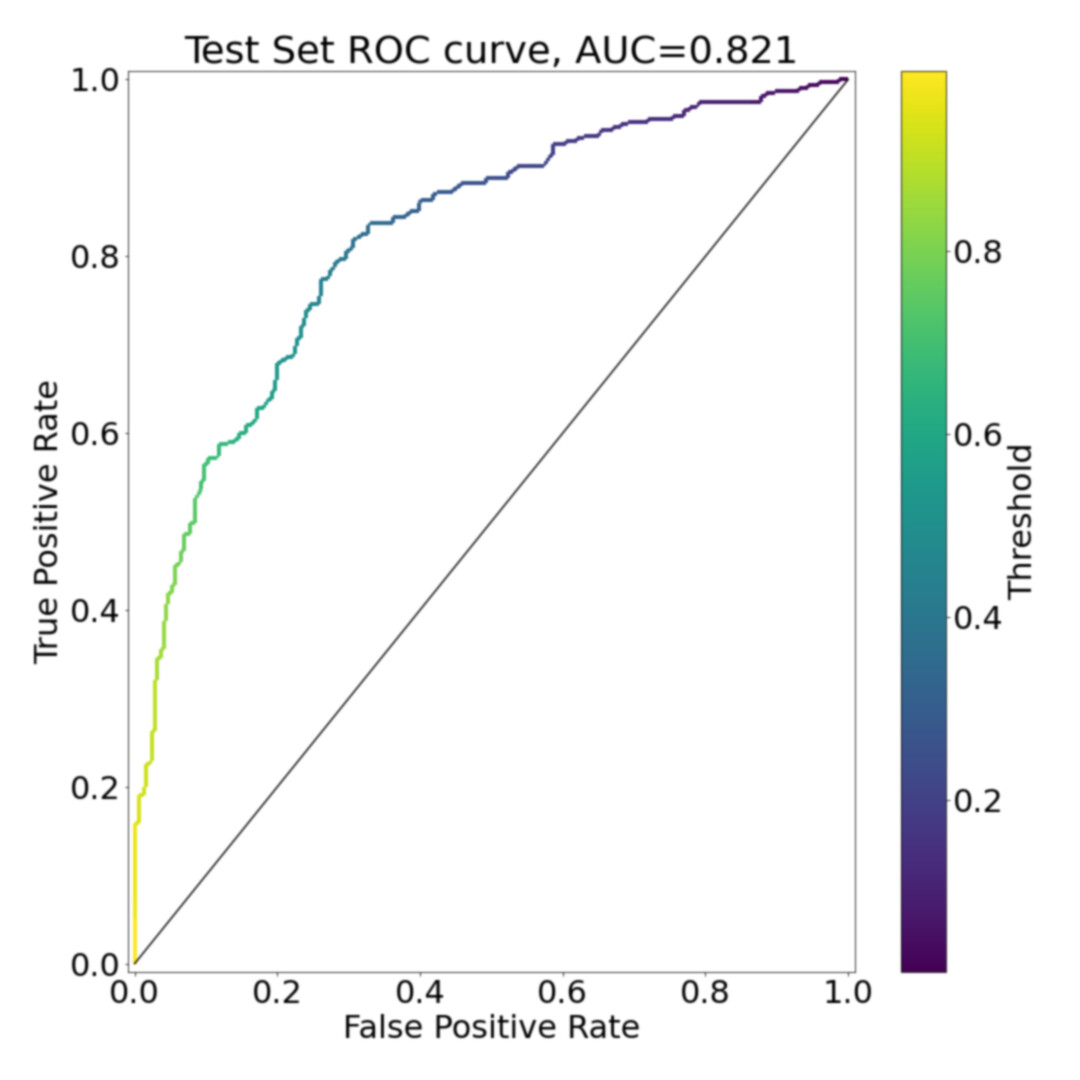}            
            \caption{Test-Set \ac{ROC} curve for \ac{XGBoost} model classifying whether a material is metallic or not.}
            \label{sfig:xgboost-metal-classifier-roc-test}
        \end{figure}

        The \ac{ROC} curves of the classifier can be found in Figure \ref{sfig:xgboost-metal-classifier-roc-train} and Figure \ref{sfig:xgboost-metal-classifier-roc-test}.
        
\begin{figure}[H]
            \centering
            \includegraphics[width=\textwidth,height=\textheight,keepaspectratio]{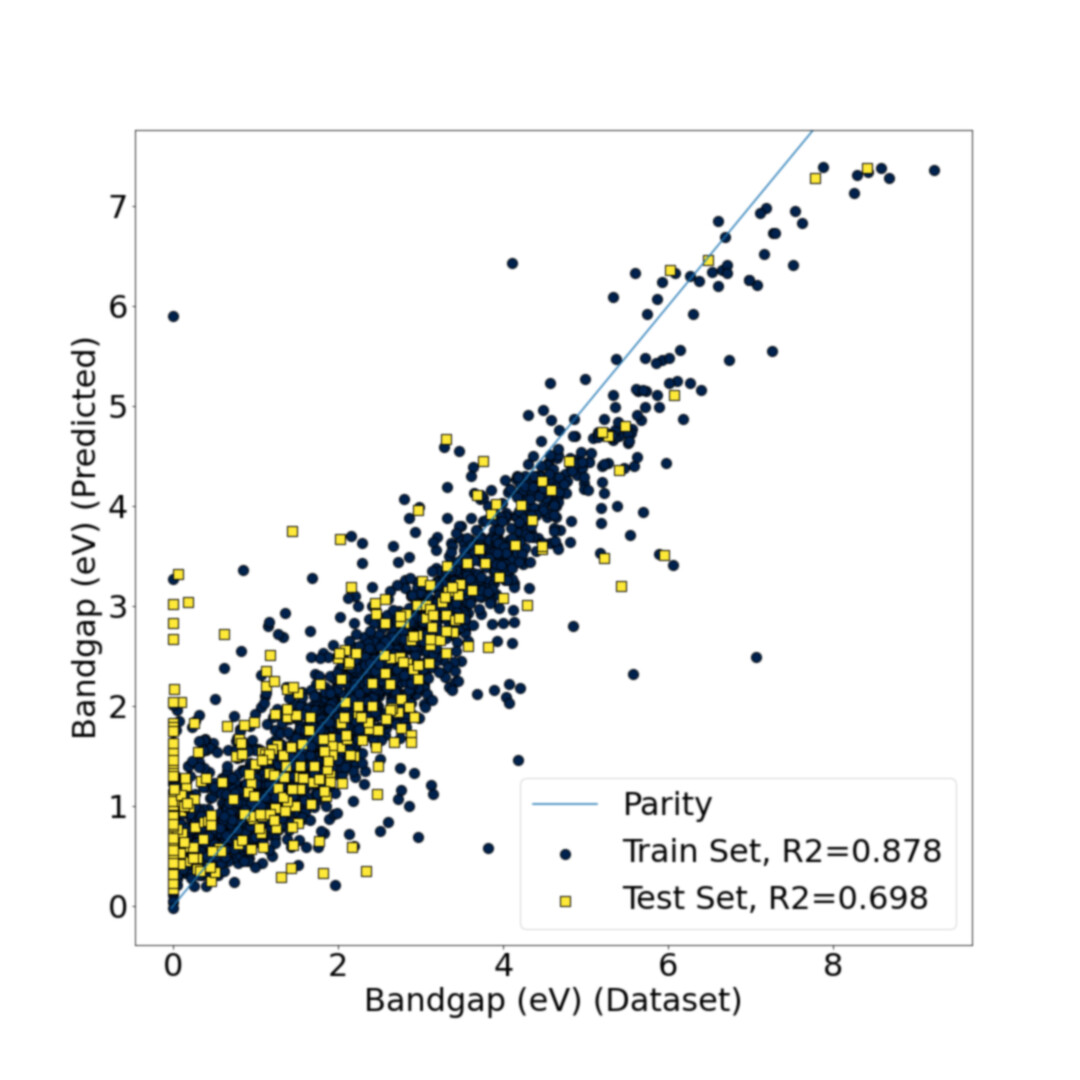}
            \caption{Parity plot showing \ac{XGBoost} regression model performance on the prediction of 2D material bandgaps.}
            \label{sfig:xgboost-metal-classifier-regression-parity}
        \end{figure}

        Finally, once we had developed an automated way to separate the dataset into metals and nonmetals, we trained an \ac{XGBoost} regression model on the systems predicted to be nonmetals by the classifier. Statistics for model performance can be found in Table \ref{stab:xgboost-metal-classifier-regression-metrics}, and a parity plot for model performance can be found in Figure \ref{sfig:xgboost-metal-classifier-regression-parity}.
        
        \begin{table}[H]
            \centering
            \rowcolors{2}{gray!25}{white}
            \caption{Error metrics for the \ac{XGBoost} regression model trained on systems predicted to be nonmetals by the \ac{XGBoost} classifier.}
            \begin{tabular}{rcc}
                \rowcolor{gray!50}
                Error Metric & Training Set & Test-Set \\
                \hline
                \acs{MAE} & 0.376 & 0.645 \\
                \acs{RMSE} & 0.530 & 0.840 \\
                Max Error & 5.893 & 3.259 \\
                R\textsuperscript{2} & 0.878 & 0.698 \\
            \end{tabular}
            \label{stab:xgboost-metal-classifier-regression-metrics}
        \end{table}
        
\begin{figure}[H]
            \centering
            \includegraphics[width=\textwidth,height=\textheight,keepaspectratio]{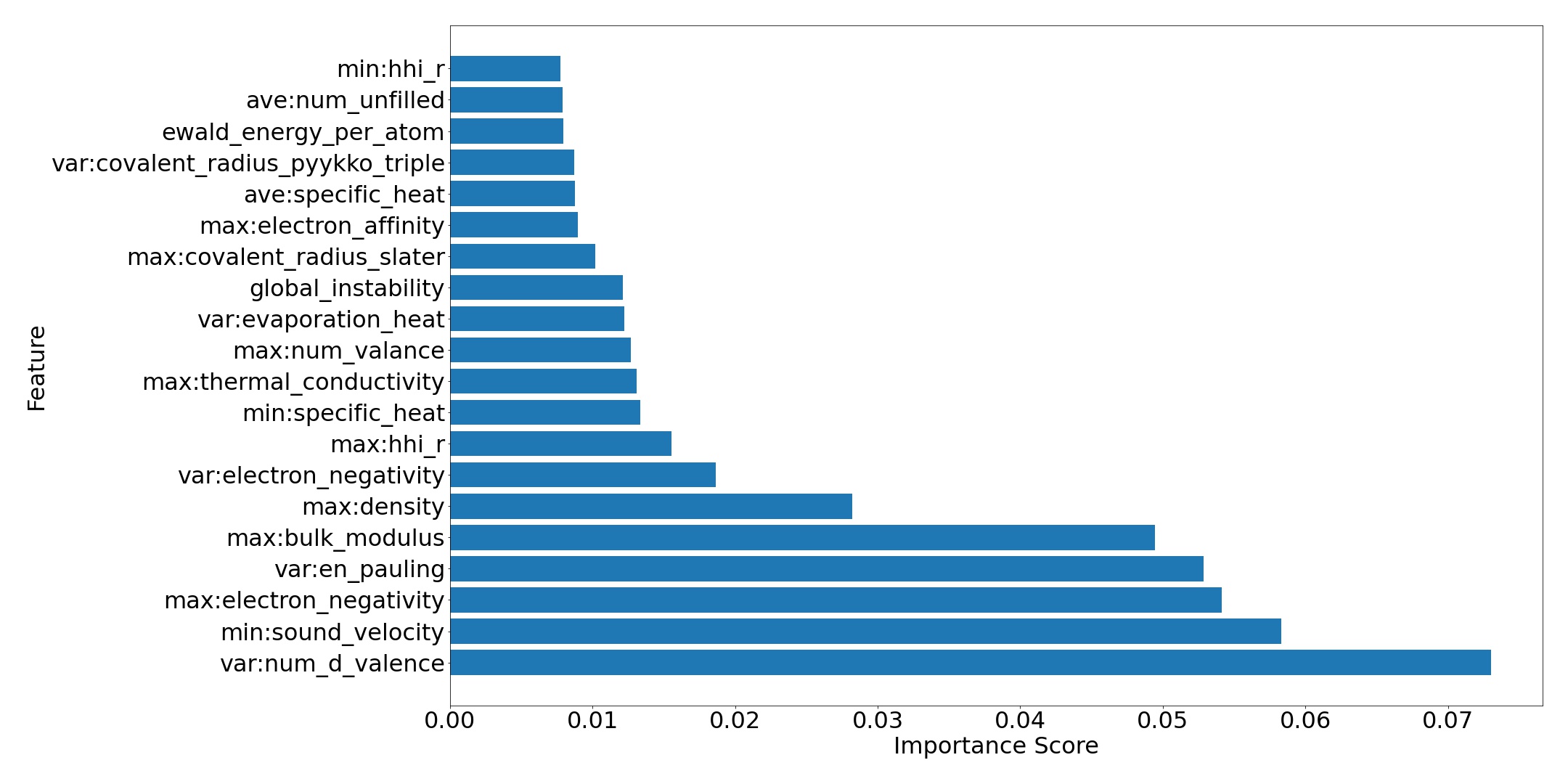}
            \caption{Importance metrics for the \ac{XGBoost} bandgap regression model on the nonmetal systems identified by the \ac{XGBoost} Metal Classifier.}
            \label{sfig:xgboost-metal-classifier-importance}
        \end{figure}

        We can also leverage the regression model to determine which features are the most important to predicting bandgap (Figure \ref{sfig:xgboost-metal-classifier-importance}).
        
\end{document}